%% file: main.tex
\documentclass[conference]{IEEEtran}
\ifCLASSINFOpdf
\usepackage[pdftex]{graphicx}
\else
\usepackage[dvips]{graphicx}
\fi
\usepackage[cmex10]{amsmath}
\DeclareMathOperator*{\argmax}{arg\,max}

\allowdisplaybreaks

\usepackage{algorithmic}
\usepackage{algorithm}

\usepackage{url}
\usepackage{colortbl}
\definecolor{Gray}{gray}{0.925}
\usepackage{multirow}
\usepackage{subfigure}
\usepackage{amssymb}
\usepackage{bm}
\usepackage[textsize=tiny, textwidth=1.4cm]{todonotes}

\usepackage[normalem]{ulem}

\usepackage[hidelinks,breaklinks]{hyperref}

\begin{document}
\setlength{\marginparwidth}{1.3cm}

\title{Finding Needles in a Moving Haystack: Prioritizing Alerts with Adversarial Reinforcement Learning}

\author{\IEEEauthorblockN{Liang Tong\IEEEauthorrefmark{1},
Aron Laszka\IEEEauthorrefmark{2},
Chao Yan\IEEEauthorrefmark{3}, 
Ning Zhang\IEEEauthorrefmark{1} and
Yevgeniy Vorobeychik\IEEEauthorrefmark{1}}
\IEEEauthorblockA{\IEEEauthorrefmark{1}Washington University in St. Louis}
\IEEEauthorblockA{\IEEEauthorrefmark{2}University of Houston}
\IEEEauthorblockA{\IEEEauthorrefmark{3}Vanderbilt University}
\IEEEauthorrefmark{1}\{liangtong, zhang.ning, yvorobeychik\}@wustl.edu, 
\IEEEauthorrefmark{2}alaszka@uh.edu, \IEEEauthorrefmark{3}chao.yan@vanderbilt.edu}

%\IEEEoverridecommandlockouts
%\makeatletter\def\@IEEEpubidpullup{6.5\baselineskip}\makeatother
%\IEEEpubid{\parbox{\columnwidth}{
%    Network and Distributed Systems Security (NDSS) Symposium 2020\\
%    23-26 February 2020, San Diego, CA, USA\\
%    ISBN 1-891562-61-4\\
%    https://dx.doi.org/10.14722/ndss.2020.23xxx\\
%    www.ndss-symposium.org
%}
%\hspace{\columnsep}\makebox[\columnwidth]{}}

\maketitle

\input{content/abstract}

\input{content/introduction}

\input{content/system_model}
\input{content/game_formulation}

\input{content/game_solver}

\input{content/case_studies}

\input{content/related_work}

\input{content/conclusion}

\bibliographystyle{IEEEtranS}
\bibliography{main}

\input{content/appendix}

\end{document}

%% file: content/abstract.tex
\begin{abstract}
Detection of malicious behavior is a fundamental problem in security.
One of the major challenges in using detection systems in practice is
in dealing with an overwhelming number of alerts that are triggered by
normal behavior (the so-called false positives), obscuring alerts
resulting from actual malicious activity.
%As a result, actual attacks are buried among the plethora of logs corresponding to normal behavior.
% 
While numerous methods for reducing the scope of this issue have been
proposed, ultimately one must still decide how to prioritize which
alerts to investigate, and most existing prioritization methods are
heuristic, for example, based on suspiciousness or priority scores.
We introduce a novel approach for computing a policy for prioritizing alerts
using adversarial reinforcement learning.
% 
% CHECK
Our approach assumes that the attacker knows the full state of the
detection system and the defender's alert prioritization policy, and
will dynamically choose an optimal attack.
%and is therefore capable of making accurate estimation of the defender policy. Using the defender state and policy estimation, the attacker will dynamically choose an optimal attack to maximize his probability of avoiding the detection. 
% capable of dynamically choosing an optimal attack as a function of this state, as well as of the alert prioritization policy.
%To defend against such threats, we propose an approach
%for robustly and dynamically prioritizing alerts.
The first step of our approach is to capture
the interaction between the defender and attacker in a game theoretic
model.
To tackle the computational complexity of solving this game to obtain a
dynamic stochastic alert prioritization policy, we
propose an adversarial reinforcement learning framework.
In this framework, we use neural reinforcement learning to compute
best response policies for both the defender and the adversary to an
arbitrary stochastic policy of the other.
We then use these in a double-oracle framework to obtain an
approximate equilibrium of the game, which in turn yields a robust
stochastic policy for the defender.
%In this framework, we alternate the following steps: a) compute an equilibrium of a
%restricted game using a small subset of policies for the attacker and
%defender, and b) use reinforcement learning to compute approximate
%best-response policies to this equilibrium for both players, and augment the game with
%the resulting optimal policies.
Extensive experiments using case studies in fraud and intrusion
detection demonstrate that our approach is effective in creating
robust alert prioritization policies.
\end{abstract}

%% file: content/introduction.tex
\section{Introduction}
\label{sec:introduction}

One of the core problems in security is \emph{detection} of malicious behavior, with examples including detection of malicious software, emails, websites, and network traffic. There is a vast literature on detection approaches, ranging from signature-based to machine-learning based~\cite{buczak2016survey,milenkoski2015evaluating,sommer2010outside}.
Despite best efforts, however, false positives are inevitable. Moreover, one cannot in general reduce the rate of false alarms without missing some real attacks as a result.
Under the pressure of practical considerations such as liability and accountability, these systems are often configured to produce a large amount of alerts in order to be sufficiently sensitive to capture most attacks.
As a consequence, cybersecurity professionals are routinely inundated with alerts, and must sift through these overwhelmingly uninteresting logs to identify alerts that should be prioritized for closer inspection.
% those using machine learning techniques such as classification and anomaly detection.
%While a common theme in detection is to limit false positives, many
%practical considerations exhibit a strong force to ensure that all
%potentially malicious behavior is logged in an alert stream.
% 
%One such force is retroactive auditing and remediation in response to
%a detected attack (for example, stolen data), where past logs are necessary to fully understand the nature and scope of compromise. 
 % and to guide deployment of mitigations as well as fulfill reporting requirements (e.g., notify people that their credit card and/or other personal information may have been compromised).

%\nznote{logging and alert display seems to be equivalent in this paper, but in real life this is different. For forensic purpose, you have have a sound method to log everything, but only display a subset of these to the incident response team. One naive way to accomplish this is to look at the most severe ones.}
%\iAron{I agree, just because we need everything for post-incident audit, does not mean that we need to raise an alert for everything. I would remove the post-incident audit argument, and focus on the difficulty of distinguishing false and true alerts without manual inspection.}

% 
A considerable literature has therefore emerged attempting to reduce
the number of false alerts without significantly affecting the ability
to detect malicious behavior~\cite{hubballi2014false,salah2013model,ho2017detecting}.
Most of these attempt to add meta-reasoning on top of detection
systems that capture broader system state, combining related alerts,
escalating priority based on correlated observations, or using alert
correlation to dismiss false alarms~\cite{vasilomanolakis2015taxonomy}.
Nevertheless,
%the fundamental tension remains that the system must
%ultimately be sensitive enough to ensure that we capture most of the attacks,
%enable \Aron{For the same reasons as above, I an not sure if ``forensic'' is what we should focus on here.} forensic investigations
%(that is, no malicious behavior should be entirely hidden from the
%logging mechanism),
despite significant advances, there are typically still vastly more alerts than time to investigate them.
With this state of affairs, alert prioritization approaches have
emerged, but rely predominantly on predefined heuristics, such as sorting
alerts by suspiciousness score or by potential associated risk~\cite{alsubhi2012fuzmet}.
However, any policy that \emph{deterministically} orders alerts potentially opens the door for determined attackers who can simply choose attacks that are rarely investigated, thereby evading detection.

%\nznote{I believe vern paxon has similar work in this domain, he highlighted in his talk of "Finding Very Damaging Needles in Very Large Haystacks" which is fairly well known to the community, we might want to briefly acknowledge that and show why it is lacking}
%\iAron{It seems that the talk (\protect\url{https://corelight.blog/2017/09/26/finding-very-damaging-needles-in-very-large-haystacks/}) is based on his USENIX paper: \protect\url{https://www.usenix.org/conference/usenixsecurity17/technical-sessions/presentation/ho} They introduce a spear-phishing detector, which (like other efforts toward detection) is mostly orthogonal to our research. In any case, here is a ref: \cite{ho2017detecting}}

Building on the observation of the fundamental tradeoff between false alert and attack detection rate, we propose a novel computational approach for robust alert prioritization to address the challenge. Our approach assumes a strong attacker who knows the full state of the detection environment including which alerts have been triggered, which have been investigated in the past, and even the defender's policy. We also assumed that the adversary is capable of finding and utilizing a near optimal attack strategy against the defender policy based on his knowledge of the system and defending policy. To defend against such a strong attacker, we propose to compute the optimal stochastic dynamic defender policy that chooses the alerts to investigate as a function of the observable state, and that is robust to our threat model.
%attacks that attempt to circumvent this policy.
At the core of our technical approach is a combination of game theory with \emph{adversarial reinforcement learning (ARL)}. Specifically, we model the problem of robust alert prioritization as a game in which the defender chooses its stochastic and dynamic policy for prioritizing alerts, while the attacker chooses which attacks to execute, also dynamically with full knowledge of the system state.
Our computational approach first uses neural reinforcement learning to compute approximately optimal policies for either player in response to a fixed stochastic policy of their counterpart.
It then uses these (approximate) best response \emph{oracles} as a part of a double-oracle framework, which iterates two steps: 1) solve a game involving a restricted set of policies by both players, and 2) augment the policy sets by calling the best response oracle for each player.
\emph{Note that our approach is completely orthogonal to methods for reducing the number of false positive alerts, such as alert correlation}, and is meant to be used in combination with these, rather than as an alternative.
In particular, we can first apply alert correlation to obtain a reduced set of alerts, and subsequently use our approach for selecting which alerts to investigate.
Since alert correlation cannot be overly aggressive in order to ensure that we still capture actual attacks, the number of alerts often still significantly exceeds the investigation budget.
%\Aron{Conventional methods do not include any prioritization? (If I am not mistaken, Suricata for example does include some priorities, to which we compare our approach.) If we have space, we could discuss how and why to combine correlation with our framework.}

We evaluate our approach experimentally in two application domains:
intrusion detection, where we use the Suricata open-source
intrusion-detection system (IDS) with a network IDS dataset, and 
fraud detection, with a detector learned from data using machine
learning.
In both settings, we show that our approach is significantly more
effective than alternatives with respect to our threat model.
Furthermore, we demonstrate that our approach remains highly effective, and better than baseline alternatives in nearly all cases, even when certain assumptions of our threat model are violated.
% , and \change{is}{remains} effective even when specific assumption of our threat model are violated.

%\iAron{Should we add ``The rest of this paper is organized as follows...'' boilerplate?}

%% file: content/system_model.tex
\section{System Model}
\label{sec:system_model}

\begin{figure*}[h]
\centering
\includegraphics[width=0.7\textwidth]{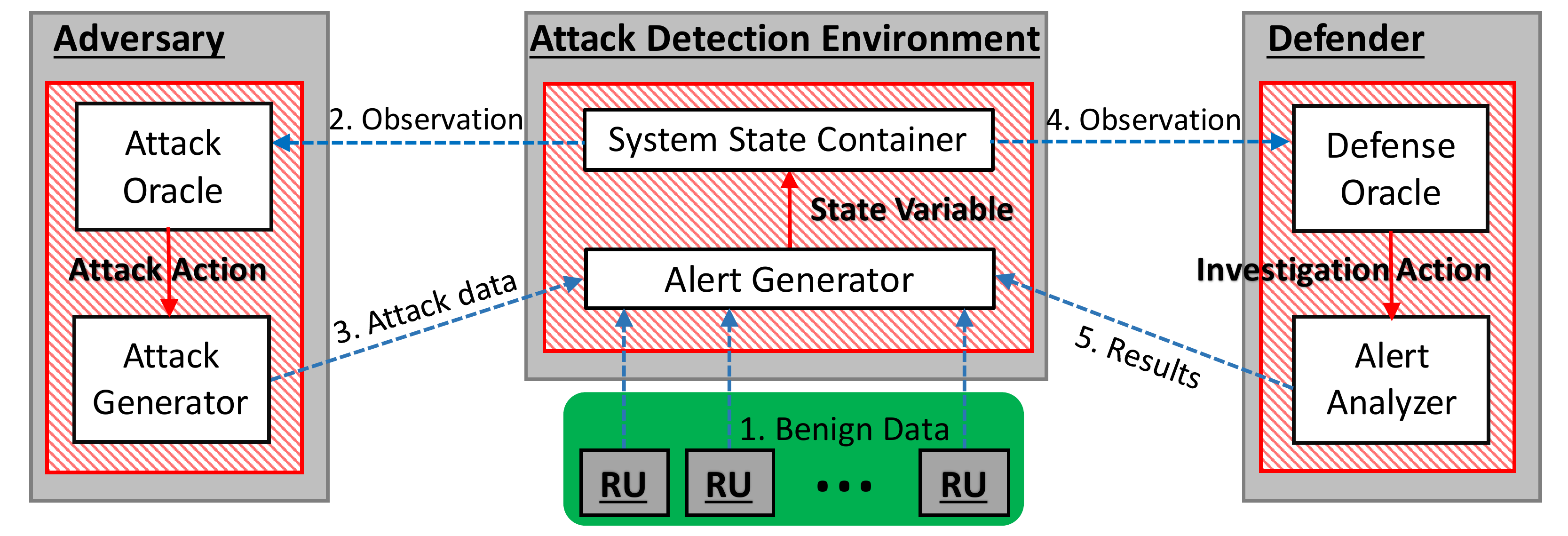}
\caption{System model.  The \emph{Attack Oracle} computes the attacker's policy for executing attacks, which is implemented by the \emph{Attack Generator} and then triggers alerts observed by the \emph{Attack Detection Environment}.  The \emph{Defense Oracle} computes the defender's alert prioritization policy, which is implemented by the \emph{Alert Analyzer}.}
%\nznote{is attack detection system the same as ADE?}

\label{fig:system_model}
\end{figure*}
%\nznote{is it possible to add orders and sequences in fig 1?}

\subsection{Overview}
As displayed in Figure \ref{fig:system_model}, our system is partitioned into four major components: a group of \emph{regular users} (RU), an \emph{adversary} (also called attacker), a \emph{defender}, and an \emph{attack detection environment} (ADE).  

%\Aron{I'm not sure if we need to model regular users explicitly, this might just raise questions}
The regular users (RU) are the authorized users of a system.
In contrast, the adversary is a sophisticated actor who attacks the target computer system. 
%\Aron{I suggest breaking this sentence up into two.}
The attack detection environment (ADE) models the combination of the software artifact that is responsible for monitoring the system (e.g., network traffic, files, emails) and raising alerts for observed suspicious behavior, as well as relevant system state.
System state includes attacks that have been executed (unknown to the defender), and alerts that have been investigated (known to both the attacker and defender).
Crucially, the alerts triggered in the ADE may correspond either to behavior of the normal users RU, or to malicious behavior (attacks) by the adversary.
%is a software that monitors users' data and generates alerts for potential policy violations and malicious activities. 
%These alerts can be triggered either by the benign data from RUs the malicious data produced by the adversary.
We divide time into a series of discrete time periods.
The defender is limited in how many alerts
%\Aron{``it''? or gender pronoun?}
it can investigate in each time period and must select a small subset of alerts for investigation, while the adversary is limited in how many attacks it executes in each time period.
%The defender is an analyst or administrator of the ADS, who develops policies to investigate the alerts generated by the ADS.
%\Aron{This is a pretty strong assumption, some might argue that this is not ``without loss of generality.''} 
%We assume that all the components of our system operate in an array of continuous time slots with a fixed length.
The full system operates as follows for a representative time period (see again the schematic in Figure~\ref{fig:system_model}):
\begin{enumerate}
\item Benign alerts are generated by the ADE.
\item These alerts, and the remaining ADE system state (such as which alerts from past time periods have not yet been investigated, but could be investigated in the future), are observed by the attacker, who executes a collection of attacks.
  %\nznote{why would the attacker observe all the alerts? Maybe we can consider saying we are assuming a very strong adversary that can even observe our internal state such as a malicious insider. And that the goal of the attacker is to remain undetected after the detect}
\item The attacks trigger new alerts.  These are arbitrarily mixed into the full collection of alerts, which is then observed by the defender.
\item The defender chooses a subset of alerts to investigate.  The ADE state is updated accordingly, and the process repeats in the next time period.
\end{enumerate}

Next, we describe our model of the alert detection environment, our threat model, and our defender model.
The full list of notation that we use in the model is presented in Table~\ref{tab:notations}.

\begin{table}[]
\caption{Notation summary.}
\label{tab:notations}
\centering
\begin{tabular}{|l|p{180pt}|}
\hline
\textbf{Notation}        & \textbf{Interpretation}                                                           \\ \hline
\multicolumn{2}{|c|}{Constants and functions}                                              \\ \hline
\rowcolor{Gray} $A$             & Types of attacks                                                         \\ \hline
$T$             & Types of alerts                                                          \\ \hline
\rowcolor{Gray} $C_t$           & Cost of investigating an alert of type $t \in T$                         \\ \hline
$B$             & Defender's budget                                                        \\ \hline
\rowcolor{Gray} $E_a$           & Cost of mounting an attack of type $a \in A$                             \\ \hline
$D$             & Adversary's budget                                                       \\ \hline
\rowcolor{Gray} $P_{a,t}(n)$    & Probability that an attack $a \in A$ raises $n$ alerts of type $t \in T$ \\ \hline
$\mathcal{F}_t$ & Probability distribution of false alerts of type $t \in T$               \\ \hline
\rowcolor{Gray} $L_a$           & Loss inflicted by an undetected attack $a \in A$                         \\ \hline
$\tau$          & Temporal discounting factor                                              \\ \hline
\multicolumn{2}{|c|}{State variables (Time slot $k \in \mathbb{N}$)}                          \\ \hline
\rowcolor{Gray} $N_t^{(k)}$     & Number of uninvestigated alerts of type $t \in T$                    \\ \hline
$M_a^{(k)}$     & Indicator of whether an attack of type $a \in A$ was mounted             \\ \hline
\rowcolor{Gray} $S_{a,t}^{(k)}$ & Number of alerts of type $t \in T$ raised due to attack $a \in A$    \\ \hline
$R^{(k)}_{+1}$       & Reward obtained by the defender                                          \\ \hline
\multicolumn{2}{|c|}{Actions, policies, and strategies}                                                             \\ \hline
%\rowcolor{Gray} $\alpha_{+1,t}$      & Number of alerts of type $t \in T$ to be investigated                \\ \hline
%$\alpha_{-1,a}$      & Probability of mounting an attack of type $a \in A$                  \\ \hline
\rowcolor{Gray} $\bm{\alpha}_{v}$      & Action of player $v \in \{-1, +1\}$                \\ \hline
$\bm{\pi}_{v}$      & Policy (i.e., pure strategy) of player $v \in \{-1, +1\}$                \\ \hline
\rowcolor{Gray} $\bm{\sigma}_{v}$      & Mixed strategy of player $v \in \{-1, +1\}$                  \\ \hline
\end{tabular}
\end{table}

\subsection{Attack Detection Environment (ADE) Model}

Our model of the attack detection environment (ADE) captures a broad array of detection settings, including credit card fraud, intrusion, and malware detection.
In this model, the ADE is composed of two parts: an \emph{alert generator} (such as an intrusion detection system, like Suricata) and \emph{system state}.

An alert generator produces a sequence of alerts in each time period.
%, which
We aggregate alerts based on a finite predefined set of types $T$.
For example, an alert type may be based on the application layer it was generated for (HTTP, DNS, etc), port number or range, destination IP address, and any other information that's informative for determining the nature and relative priority of alerts.
We can also define alert types for meaningful sequences of alerts.
Indeed, the notion of alert types is entirely without loss of generality---we can define each type to be a unique sequence of alerts, for example---but in practice it is useful (indeed, crucial for scalability) to aggregate semantically similar alerts.

At the end of each time period the system generates a collection of alert \emph{counts} for each alert type $t \in T$.
%While this may appear to ignore potentially relevant information about sequences,
We assume that normal or benign behavior generates alerts according to a known distribution $\mathcal{F}$, where $\mathcal{F}_t(n)$ is the marginal probability that $n$ alerts of type $t$ are generated.
We also refer to this as the distribution of \emph{false alarms}, since if the defender were omniscient, they would never trigger such alerts.
Note that in practice it is not difficult to obtain the distribution $\mathcal{F}$.
Specifically, we can use past logs of \emph{all} alerts over some time period to learn the distribution $\mathcal{F}$.
Since the vast majority of alerts in real systems are in fact false positives, any unidentified true positives in the logs will have a negligible impact.\footnote{If we are concerned about these poisoning the data, we can use robust estimation approaches to mitigate the issue~\cite{Vorobeychik18book}.}
%The distribution $\mathcal{F}$ can be learned from past alerts, the predominant majority of which are normal.

%In each time period, the alert generator produces a number of alerts, which are partitioned into a set of alert types $T$.
%In addition to the true alerts, the alert generator also generates \emph{false alerts}, which are triggered by benign data from regular users.
%We assume that the number of false alerts of type $t$ follows some known probability distribution $\mathcal{F}_t(n)$, measured by the probaility of generating $n$ false alerts of type $t$ at each time slot.

%Afterward,  all the raised alerts are rendered to the defender for further inspection. 
%Based on the investigation results returned, the alert generator sends corresponding \emph{state variables} of the current time slot to the global state container. 
%\Aron{Now I see what you meant earlier. However, these should not be a part of the \emph{detection system}.}
We use three matrices to represent the state of ADE at time period $k$.
%\Aron{I suggest removing this first list, and just introducing the variables one-by-one (as you do in the following sentences).}
The first represents the counts of alerts not yet investigated, grouped by type. Formally, we denote this structure by $\mathbf{N}^{(k)} = \{N_t^{(k)}\}_{t \in T}$, where $N_t^{(k)}$ is the number of alerts of type $t \in T$ that were raised but have not been investigated by the defender.
This is observed by \emph{both} the defender and the attacker.
The second describes which attacks have been executed by the adversary; formally, $\mathbf{M}^{(k)}=\{M_a^{(k)}\}_{a \in A}$, where $M_a^{(k)}$ is a binary indicator where $M_a^{(k)}=1$ iff the attack $a$ was executed.
This matrix is only observed by the attacker.
%is an identifier that shows whether an attack of type $a \in A$ was mounted by the adversary.
%If the adversary produces an attack of type $a$, then $M_a^{(k)}=1$; otherwise, $M_a^{(k)}=0$;
Finally, we represent which alerts are raised specifically due to each attack.
Formally, $\mathbf{S}^{(k)}=\{S_{a,t}^{(k)}\}_{a \in A, t \in T}$, where $S_{a,t}^{(k)}$ represents the number of alerts of type $t \in T$ raised due to attack $a$.
This is also only observed by the attacker.
%These variables can be partially or entirely observed by the defender and adversary.
%In addition, they are applied to determine the rewards of the adversary and defender for the current time slot, as a measure of performance. 

\subsection{Threat Model}

{\bf Adversary's Knowledge}.
We consider a strong attacker who is capable of observing the current state of the ADE.
This obviates the need to make specific (and potentially erroneous) assumptions about information actually available to the attacker about system state; in practice, given the zero-sum nature of the encounter we consider below, having a less informed attacker will only improve the defender's utility.
Additionally, the attacker knows the randomized \emph{policy} used by the defender for choosing which alerts to inspect (more on this below), and inspection decisions in previous rounds, but not the inspection decision in the current round (which happens after the attack).

%In practice, the adversary can be either an external attacker or an insider who is capable of accessing global states of the ADS.
%\Aron{What do you mean by ``exploiting the global states of the ADS''?}
%\Aron{We need to explain that this is how we model the attacker, we do not assume that attackers have this exact architecture in practice.}
%The adversary is logically devided into two parts:  an attack oracle, and an attack generator.\footnote{This is our model of the adversary; we do not assume that it has this exact architecture in practise.}
%The attack oracle is applied to develop attack policies through the observations of the adversary and its rewards obtained.
%The attack generator is used to produce actual malicious data based on the attack policies, subject to the adversary's attack budget $D$. 

%Next, we provide detailed descriptions of the adversary's capability and goal.

{\bf Adversary's Capabilities}.
In each time period, the adversary can execute multiple actions $a$ from a set of possible (representative) actions $A$.\footnote{In practice, actions in $A$ correspond to equivalence classes of attacks; for example, $a \in A$ could be a representative denial-of-service attack.}
Each attack action $a \in A$ stochastically triggers alerts according to the probability distribution $P$, where $P_{a, t}(n)$ is the marginal probability that action $a$ generates $n$ alerts of type $t$.
These probabilities can be learned by replaying known attack actions through actual detectors (as we do in the experiments below), ideally as a part of a full dataset which includes a mix of benign and malicious behavior.
Commonly, alerts are generated deterministically for given attack actions; it is evident that our model admits this as a special case (i.e., $P_{a, t} \in \{0, 1\}$).
However, our generality allows us to handle important cases where alerts are, indeed, stochastic.
For example, consider a \textit{Port Scan} attack (as a part of a reconnaissance step).
Port scan alert rules commonly consider the number of certain kinds of packets (such as ICMP packets) observed over a small time period (say, several seconds), and raise an alert if this number exceeds a predefined threshold.
The number of such packets, of course, also depends on background traffic, which is stochastic, so that the triggering of the alert is also stochastic if the attack is sufficiently stealthy to avoid exceeding such a threshold in isolation.
%, but our model is capable of capturing a broad array of realistic situations.
%In our model, the adversary is assumed to be capable of producing \emph{worst-case} attacks:
%The adversary has full knowledge of the global states of the ADS when mounting attacks in each time slot $k \in \mathbb{N}$.
%In practise, such attackers can be insiders which are authorized to access the target system\cite{mathew2010,chen2012}, \Aron{For a security venue, we will need a better argument than referring to the game theory literature (since some reviewers may be skeptical of that line of work)} or outsiders which can illegally eavesdrop the ADS.
% and is widely applied in previous works of  game-theoretical models for alert investigation problems \cite{sinha2018, yan2018, laszka2017}.

% These alerts can be triggered by multiple types of malicious data produced by the adversary (henceforth, \emph{true alerts}).

%Let $A$ be the set of attack types of the attacker. 
%We assume that for any attack $a \in A$, it triggers $n$ alerts of type $t \in T$ with probability $P_{a, t}(n)$.

Let $E_a$ be the cost for executing an attack $a \in A$.
One method to estimate these costs is to examine the difficulty of executing the exploit based on the CVSS complexity metrics.
The main limitation to the attacker capabilities is a budget constraint $D$ that limits how many, and which combination of, attacks can be executed.\footnote{Note that this easily admits the possibility of multiple attackers, where $D$ becomes the total budget of all attackers.  This case is equivalent to assuming that attackers coordinate.  This is a safe assumption, since if they do not, the defender's utility can only increase.}
While it is difficult to reliably estimate this budget, our case studies in Section~\ref{sec:case_studies} demonstrate that our approach is robust to uncertainty about this parameter.
%\iAron{Here, we should (1) acknowledge that in practice it is difficult to estimate this budget and (2) highlight that our approach is robust against uncertainty about the attacker's budget, which we demonstrate experimentally (reference results from case study)!}
Specifically, any attack decision $\bm{\alpha}_{-1}$ with $\alpha_{-1,a}$ the probability that the attack $a$ is executed by the attacker 
%any subset $\tilde{A} \subseteq A$ of attacks chosen by the attacker
in a given time period, must abide by the following constraint:
%\Aron{I think that it might better to discuss first the adversary's capabilities and then its goal since the goal formulation seems to depend on capabilities (e.g., budget).}
%As the adversary incurs a cost $E_a$ for mounting attack of type $a \in A$, and the adversary has a budget $D$ in each time slot, the expected cost of the attacks mounted is bounded as displayed below: 
\begin{equation}
\sum_{a \in A} \alpha_{-1,a} E_a \leq D.
\label{eq:adv_capability}
\end{equation} 
%\iAron{Here, we use $\tilde{A}$ to denote an attack. This notation does not seem to appear anywhere else in the paper, and it is not listed in the table of symbols either. It might be better to use the $\bm{\alpha}$ notation (as we do later).}

For our purposes, it is useful to represent the attacker as consisting of two modules: \emph{Attack Oracle} and \emph{Attack Generator}, as seen in Figure~\ref{fig:system_model}.
The attack oracle runs a \emph{policy}, which maps observed the state of the ADE to attacks that are executed.
In each time period, after observing ADE state, the attack oracle chooses attack actions, which are then executed by the attack generator, triggering alerts and thereby modifying the state of the ADE.
Below we present our approach for approximating the optimal attack policies.

{\bf Adversary's Goals}.
The adversary aims to successfully execute attacks.
Success entails avoiding being detection by the defender, \emph{which only happens if alerts associated with an attack are inspected}.
Thus, if an attack triggers a collection of alerts, but none of these are chosen by the defender to be inspected in the current round,
%within a reasonable time period \nznote{within the next $N$ rounds},
the attack succeeds. 
Different attacks, however, entail different consequences and, therefore, different rewards to the attacker (and loss to the defender).
As a result, the adversary will ultimately need to balance rewards to be gained from successful attacks and the likelihood of being detected.
%The goal of the adversary is to produce attacks that are not detected by the defender.
%\Aron{Again, should we restrict attackers to one with economic motivations?}
%As described above, the attacks can be of various types with diverse utilities.
%Meanwhile, different types of attacks can exhibit different probabilities of triggering various alerts.
%To achieve its goal, the adversary develops attack strategies to allocate its budget among various types of attacks and optimize the tradeoff between reward and risk.
%\Aron{This is the first time that a budget is mentioned, it should be introduced before.}

\subsection{Defender Model}
%The defender is divided into two components: a \emph{defense oracle} and an \emph{alert analyzer}.
%The defense oracle is a decision-making system that  observes states of the ADS and develops its investigation strategies.
%The defense oracle is a decision-making system that develops an investigation policy,
%\Aron{It will be important to note that the defender cannot observe the complete state (e.g., cannot observe $\mathbf{M}$)}
%which is applied by the alert analyzer to conduct inspections on a subset of the alerts forwarded by the alert generator of the ADS, subject to a budget $D$.
%The alert analyzer is a transparent entity such that it has no knowledge about whether an alert to being investigated is triggered by benign data from regular users, or attacks mounted by the adversary.
%Once the inspection is completed, the analysis results are forwarded back to the ADS.

%Next, we describe the capability and goal of the defender in detail. 

{\bf Defender's Knowledge}.
Unlike the adversary, the defender can only partially observe the state of the ADE.
In particular, the defender only observes $\mathbf{N}^{(k)}$, the numbers of remaining uninvestigated alerts, grouped by alert type (since clearly the defender cannot directly observe actually attacks).
%In addition, we assume that the defender knows $A$, the set of possible attacks (but, of course, not the \emph{actual} attacks executed by the attacker).
%We now observe that this assumption is far less limiting than it may first appear.
%In particular, rather than viewing $A$ as the set of attacks, we can view it as the set of \emph{equivalence classes} of attacks, where two attacks $a,a'$ are approximately equivalent if the distributions of alerts $P_a$ and $P_{a'}$ are similar (e.g., in terms of KL-divergence), and they have similar consequences.
%For our purposes, therefore, the defender only needs to know \emph{representative} attacks from each equivalence class, and we view the set $A$ as the set of such representative attacks (thus, we would in practice use examples of the \emph{kinds} of attacks one is concerned about for the purposes of computing a robust alert prioritization policy).
In addition, we assume that the defender knows the attack budget and costs of (representative) attacks.
In our experiments, we study the impact of relaxing this assumption (see Sections \ref{sec:case_study_fraud} and \ref{sec:case_study_ids}), and provide practical guidance on this issue.

{\bf Defender's Capabilities}.
The defender chooses subsets of alerts in $\mathbf{N}^{(k)}$ to investigate in each time period $k$.
This choice is constrained by the defender's budget, which in practice can translate to time the defender has to investigate alerts.
Since different types of alerts may need different amounts of time to investigate, or more generally, incur varying investigation costs, the budget constraint is on the total cost of investigating chosen alerts.
Formally, let $C_t$ be the investigation cost of an alert of type $t$, and let $\alpha_{+1,t}^{(k)}$ be the number of alerts of type $t$ chosen to be investigated by the defender in period $k$.
Then the budget constraint takes the following mathematical form:
\begin{equation}
\sum_{t \in T} C_{t} \alpha_{+1,t}^{(k)} \leq B.
\label{eq:def_capability1}
\end{equation}
An additional constraint imposed by the problem definition is that the defender can only investigate existing alerts:
%Meanwhile, the number of alerts to be investigated for each type should be upper bounded by the actual number of alerts raised for that type, as presented below:
\begin{equation}
\forall t \in T: \alpha_{+1,t}^{(k)} \leq N_t^{(k)}.
\label{eq:def_capability2}
\end{equation}

Just as with the adversary, it is useful to represent the defender as consisting of two modules: \emph{Defense Oracle} and \emph{Alert Analyzer}, as shown in Figure~\ref{fig:system_model}.
The defense oracle runs a \emph{policy}, which maps \emph{partially observed} state of the ADE to the choice of a subset of alerts to be investigated.
In each time period, after observing the set of as yet uninvestigated alerts, the defense oracle chooses which alerts to investigate, and this policy is then implemented by the alert analyzer, which thereby modifies ADE state (marking the selected alerts as having been investigated).
Below we present our approach for approximately computing optimal defense policies that are robust to attacks as defined in our threat model above.

{\bf Defender's Goals}.
The goal of the defender is to guard a computer system or network by detecting attacks through alert inspection. 
%As described above, the attacks and alerts are in diverse, and the alert analyzer is transparent to the alerts.
%Furthermore, due to limited resources and overwhelming alert volume, it is infeasible for the defender to inspect all the raised alerts. 
To achieve its goal, the defender develops an investigation policy to allocate its limited budget to investigation activities in order to minimize consequences of successful attacks, where we assume that an attack will fail to accomplish its primary objectives if the alerts it causes the ADE to emit are investigated in a timely manner.

%over various types of alerts so as to maximize the possibility that true alerts are inspected.
%\Aron{Again, it might better to first discuss capabilities and then goals.}

\subsection{An Illustrative Example}

Since our system is built on top of an abstracted model of alert investigation, the results are generally applicable to a wide range of real-world problems. We will use intrusion detection as an illustrative example in this section. \textit{Port Scan} reconnaissance attack is one of the most common initial steps in remote exploitation and is a common occurrence faced by many enterprise IT professionals. In a Suricata IDS system, each alert item has different levels of categorization. For example, at the lowest layer, the port scan may trigger two types of alert, 1) \textit{Httprecon Web Server Fingerprint Scan}, and 2) \textit{ET SCAN NMAP -sO}. At a higher level, these alerts can be categorized into \textit{attempted-recon} (since both reflect potential reconnaissance efforts by the attacker), as is the case in the \emph{Emerging Threats Ruleset} of Suricata. A defender can choose different granularities of attack categorization to map the IDS alert types into the abstracted types in our proposed model based on individual needs. Besides categorization, the defender can also make use of other attributes in the IDS alerts to aid in abstracted type assignment. For example, a port scan on the enterprise file server can be assigned to the abstracted type of \textit{high-risk-recon}, while a port scan on employee desktop can be assigned to \textit{attempted-recon}. 

In addition to the alerts corresponding to an actual attack action, normal user behavior can generate false positive alerts. For example, a user who is scraping the web for weather data monitoring may trigger the \textit{ET POLICY POSSIBLE Web Crawl using Curl}, which is grouped into the \textit{attempted-recon} type by the same \emph{Emerging Threats} Suricata ruleset. Leveraging the proposed game-theoretic model on these abstracted alerts, it is possible for the defender to devise an optimal defense policy for a wide range of alert applications even in the face of possible false positives. 

%% file: content/game_formulation.tex
\section{Game Theoretic Model of\\Robust Alert Prioritization}
\label{sec:game_formulation}

We now turn to the proposed approach for robust alert prioritization.
We model the interaction between the defender and attacker as a \emph{zero-sum} game, which allows us to define and subsequently compute robust stochastic inspection policies for the defender.
In this section, we formally describe the game model.
We then present the computational approach for solving it in Section~\ref{sec:game_solver}.
%, where each player has an inverse utility of the other and aims at maximizing its utility by strategically making a once-for-all decision from a collection of possible policies.
%Our proposed game consists of four parts: players, strategies, utilities and solution concept.
%The players are competitors in the game.
%The strategies are a plan of policy to accomplish the players' goals.
%The utilities are payoffs of the players given their policies.
%The solution concept is a formal rule that describes which policy will be adopted by each player as a result of the game. 

%Below, we formally describe the proposed game in detail.
%We present our game solution in Section~\ref{sec:game_solver}.

%\subsection{Players}
The game has two players: the defender (denoted by $v=+1$) and the adversary (denoted by $v=-1$).
Each player's strategies are policies, that is, mappings from an observed ADE state to the probability distribution over actions to take in that state.
In a given state, the defender chooses a subset of alerts to investigate; thus, the defender's set of possible actions is the set of all alert subsets that satisfy the constraints~\eqref{eq:def_capability1} and \eqref{eq:def_capability2}.
The attacker's choices in a given state correspond to subsets of actions $A$ to take.
Consequently, the set of adversary's actions is the set of all subsets of attacks satisfying constraint~\eqref{eq:adv_capability}.
Note that the combinatorial nature of both players' action spaces and of the state space makes even \emph{representing deterministic policies} non-trivial; we will deal with this issue in Section~\ref{sec:game_solver}.
Moreover, we will consider stochastic policies.
An equivalent way to represent stochastic policies is as probability distributions over deterministic policies, which map observed state to a \emph{particular} action (subset of alerts for the defender, subset of attacks for the adversary).
%Our key hypothesis is that it suffices to consider distributions which only use a small set of deterministic policies (i.e., distributions with a small support); this is borne out in our experiments.
%, as our approach below indeed returns stochastic policies with small supports in all cases.
Henceforth, we call deterministic policies of the players their \emph{pure strategies} and stochastic policies are termed \emph{mixed strategies}, following standard terminology in game theory.\footnote{At decision time, players can sample from their respective mixed strategies in each round, thereby determining their decisions in that round.  We assume that while the defender's mixed strategy is known to the attacker, the realizations, or samples, of deterministic policies drawn in each round are not observed by the attacker; for example, the sampling process can take place after the entire set of alerts in that round are observed.  Note that if we resample independently in each round, the attacker learns no additional information about the defender's policy from past rounds.}
%act as players that satisfy the following assumptions:
%\begin{enumerate}
%\item Players have complete information of others. 
%Specifically, each player has knowledge of the budget and set of policies of its component.   
%This is a strong assumption, and we relax it in our experimental evaluation (Section \ref{sec:case_study_fraud} and \ref{sec:case_study_ids}), providing guidance on how to deal with uncertainty about adversary's information.
%\item Players are rational, and they believe that their opponents are rational as well.
%Specifically, each player aims at finding its optimal policy while taking the policies which are possibly adopted by its opponent into account.
%\item Each player independently chooses an action.
%By independently, we mean that a player makes its decision without observing the policy taken by its counterpart.
%This is a common observation in practice.
%For example, when a hacker performs attacks to a network, he may have no knowledge of the action taken by the network guardian, though he knows the collection of possible behaviors of the latter, and vice versa.
%\end{enumerate}

%\subsection{Strategies}
%We assume that each player in the game executes stochastic policies by playing \emph{mixed strategies}.
%A mixed strategy is a distribution over the set of policies of a player.
%Initially, each player chooses its policy through sampling with its mixed strategy, it then sticks to this policy throughout the game.

Let $\bm{\pi}_{-1}$ denote the attacker's policy, which maps the fully observed state of ADE, $\bm{O}_{-1}^{(k)}=\langle\mathbf{N}^{(k)}, \mathbf{M}^{(k)}, \mathbf{S}^{(k)}\rangle$, to a subset of attacks.
Let
%\Aron{I think that we need to say what $\bm{\alpha}$ is (I mean beyond the mathematical definition). If we use this notation instead of $\tilde{A}$ earlier, then we can just refer back.}
$\bm{\alpha}_{-1}^{(k)} = \bm{\pi}_{-1}(\bm{O}_{-1}^{(k)})$, where $\bm{\alpha}_{-1}^{(k)}=\{\alpha_{-1,a}^{(k)}\}_{a \in A}$ are (for the moment) binary indicators with $\alpha_{-1,a}^{(k)}=1$ iff an action $a \in A$ is chosen by the attacker.
In other words, the vector $\bm{\alpha}_{-1}^{(k)}$ represents the choice of actions made by the adversary.
Similarly, $\bm{\pi}_{+1}$ denotes the defender's policy, which maps the portion of ADE state $\bm{O}_{+1}^{(k)}=\mathbf{N}^{(k)}$ observed by the defender to the number of alerts of each type to investigate.
Aalogous to the attacker, $\bm{\alpha}_{+1}^{(k)} = \bm{\pi}_{+1}(\bm{O}_{+1}^{(k)})$, where $\bm{\alpha}_{+1}^{(k)}=\{\alpha_{+1,t}^{(k)}\}_{t \in T}$ are the counts of alerts chosen to be investigated for each type $t$.
Now, notice that all alerts of type $t$ are equivalent by definition; consequently, it makes no difference to the defender which of these are chosen, and we therefore choose the fraction $\frac{\alpha_{+1,t}^{(k)}}{N_t^{(k)}}$ of alerts of type $t$ uniformly at random.

%Based on the entire observations of the system states and rewards, the attack oracle then develops its strategy: a policy $\bm{\pi}_{-1}$  that maps its observation $\bm{O}_{-1}^{(k)}=\langle\mathbf{N}^{(k)}, \mathbf{M}^{(k)}, \mathbf{S}^{(k)}\rangle$ into its attack action, which is the sequence of the probability of mounting each type of attack.
%Let $\bm{\alpha}_{-1}^{(k)} = \bm{\pi}_{-1}(\bm{O}_{-1}^{(k)})$, where $\bm{\alpha}_{-1}^{(k)}=\{\alpha_{-1,a}^{(k)}\}_{a \in A}$.
%$\{\alpha_{-1,a}^{(k)}\}_{a \in A}$ is then forwarded to the attack generator
%to produce actual attacks.

%The defender applies its strategy, a policy $\bm{\pi}_{+1}$ that maps its observation $\bm{O}_{+1}^{(k)}=\mathbf{N}^{(k)}$ into its defensaction, which is the number of each type of alerts to be investigated. 
%Let $\bm{\alpha}_{+1}^{(k)} = \bm{\pi}_{+1}(\bm{O}^{(k)})$, where $\bm{\alpha}_{+1}^{(k)}=\{\alpha_{+1,t}^{(k)}\}_{t \in T}$.
%$\{\alpha_{+1,t}^{(k)}\}_{t \in T}$ is then forwarded to the alert analyzer to allocate the investigation budget.
%That is, for the alerts of type $t \in T$, a portion of $\frac{\alpha_{+1,t}^{(k)}}{N_t^{(k)}}$ is randomly selected to be inspected.

Let $\bm{\Pi}_{v}$ be player $v$'s set of pure strategies, where each pure strategy $\bm{\pi}_v \in \bm{\Pi}_{v}$ is a policy as defined above.
%When $v=+1$, $\bm{\Pi}_{v}$ represents the defender's set of pollicies; when $v=-1$, $\bm{\Pi}_{v}$ is the set of policies of the adversary.
A mixed strategy of player $v$ is then a probability distribution $\bm{\sigma}_v = \{\sigma_v(\bm{\pi}_v)\} _{\bm{\pi}_v\in\bm{\Pi}_v}$ over the player's pure strategies $\bm{\Pi}_{v}$ where $\sigma_v(\bm{\pi}_v)$ is the probability that player $v$ uses policy $\bm{\pi}_v$.
Since a mixed strategy $\bm{\sigma}_v$ is a distribution over a finite set of pure strategies, it satisfies
$0 \leq \sigma_v(\bm{\pi}_v) \leq 1$ and $\sum_{\bm{\pi}_v \in \bm{\Pi}_v} \sigma_v(\bm{\pi}_v)=1 $.
Let $\bm{\Sigma}_v$ denote the set of all mixed strategies of player $v$.

%Let $\bm{\pi}_{+1} = \{\bm{\pi}_i\}_{i=1}^{m}$ be the defender's finite set of actions, where each action $\bm{\delta}_i$ is an investigation policy of the defender.
%A mixed strategy of the defender is a probability distribution $\bm{p}_d =\{p_d(\bm{\delta}_i)\}_{i=1}^m$ over $\bm{\Delta}$ where $p_d(\bm{\delta}_i)$ is the probabililty that the defender plays $\bm{\delta}_i$. 
%Since $\bm{p}_d$ is a distribution over a finite set, $0 \leq p_d(\bm{\delta}_i) \leq 1$ and $\sum_{i=1}^m p_d(\bm{\delta}_i)=1$.
%Similarly, let $\bm{\Phi} = \{\bm{\phi}_j\}_{j=1}^{n}$ be the adversary's finite collection of actions, where each action is one of its attack policies.
%A mixed strategy of the adversary is a probability distribution $\bm{p}_a =\{p_d(\bm{\phi}_j)\}_{j=1}^n$ over $\bm{\Phi}$ where $\bm{p}_a$ satisfies $0 \leq p_a(\bm{\phi}_j) \leq 1$ and $\sum_{j=1}^n p_a(\bm{\phi}_j)=1$. 

%\subsection{Utilities}

For any strategy profile of the two players, $(\bm{\pi}_{v}, \bm{\pi}_{-v})$, we denote the utility of each player $v$ by $U_v(\bm{\pi}_v, \bm{\pi}_{-v})$, $v\in\{+1,-1\}$.
%\footnote{Here the position of $v$ and $-v$ are interchangeable such that $U_v(\bm{\pi}_v, \bm{\pi}_{-v}) = U_v(\bm{\pi}_{-v}, \bm{\pi}_v)$.}
Since our game is \emph{zero-sum},
%(i.e.,
% \Aron{Do we talk about the zero sum assumption earlier? For someone unfamiliar with game theory for security, this reasoning might not be obvious.}
%(we are ultimately interested in robust policies for the defender),
$\sum_{v\in\{+1,-1\}} U_v(\bm{\pi}_v, \bm{\pi}_{-v}) = 0$.
When player $v$ chooses pure strategy $\bm{\pi}_v\in\bm{\Pi}_v$ and its opponent $-v$ plays mixed strategy $\bm{\sigma}_{-v}\in\bm{\Sigma}_{-v}$, then the expected utility of $v$ is
\begin{equation}
U_v(\bm{\pi}_v, \bm{\sigma}_{-v}) = \sum_{\bm{\pi}_{-v}\in\bm{\Pi}_{-v}}\sigma_{-v}(\bm{\pi}_{-v}) U_v(\bm{\pi}_v, \bm{\pi}_{-v}).
\label{eq:utility_pure_mix}
\end{equation}
Similarly, the expected utility of player $v$ when it chooses the mixed strategy $\bm{\sigma}_{v}\in\bm{\Sigma}_v$ and its opponent play the mixed strategy $\bm{\sigma}_{-v}\in\bm{\Sigma}_{-v}$ is
\begin{equation}
%\begin{split}
U_v(\bm{\sigma}_{v}, \bm{\sigma}_{-v}) = \sum_{\bm{\pi}_v\in\bm{\Pi}_v}\sigma_{-v}(\bm{\pi}_{v}) U_v(\bm{\pi}_v, \bm{\sigma}_{-v}).
%&= \sum_{\bm{\pi}_v\in\bm{\Pi}_v}\sum_{\bm{\pi}_{-v}\in\bm{\Pi}_{-v}} \sigma_{v}(\bm{\pi}_{v})\sigma_{-v}(\bm{\pi}_{-v}) U_v(\bm{\pi}_v, \bm{\pi}_{-v})
%\end{split}
\label{eq:utility_mix_mix}
\end{equation} 

Next, we describe how to compute the utility of player $v$, $U_v(\bm{\pi}_v, \bm{\pi}_{-v})$, when its policy is $\bm{\pi}_v$ and the opponent's policy $\bm{\pi}_{-v}$ are given.

Consider arbitrary pure strategies of both players, $\bm{\pi}_{+1}$ and $\bm{\pi}_{-1}$. The game begins with an initial system state $\langle \bm{N}^{(0)}, \bm{M}^{(0)}, \bm{S}^{(0)} \rangle = \langle \bm{0}, \bm{0}, \bm{0} \rangle$.
The system state is then updated in each time period $k$ as follows:

\begin{enumerate}

\item \emph{Alert investigation}. 
The defender first investigates a subset of alerts produced thus far.
Specifically, the defender chooses the number of alerts of each type to investigate $\{\alpha_{+1,t}^{(k)}\}_{t\in T}$ according to its policy $\bm{\pi}_{+1}(\bm{O}_{+1}^{(k)})$ given current observed state $\bm{O}_{+1}^{(k)}$.
%mapped from its observation $\bm{O}_{+1}^{(k)}$ through its policy $\bm{\pi}_{+1}$.
For each attack $a \in A$, let $\widetilde{M}_a^{(k)}$ be an indicator of whether attack $a$ has been executed by the beginning of time period $k$, but has not been investigated.
If $M_a^{(k)}=0$, we have $\widetilde{M}_a^{(k)}=0$ as no attack $a \in A$ has been executed.
If $M_a^{(k)}=1$, then $\widetilde{M}_a^{(k)}=1$ with probability
\begin{equation}
p_{a}^{(k)} = \prod_{t \in T} \left\lbrace  \frac{C(N^{(k)}_{t}-S^{(k)}_{a,t}, \alpha_{+1,t}^{(k)})}{C(N^{(k)}_{t}, \alpha_{+1,t}^{(k)})} \right\rbrace,
\label{eq:prob_no_inves} 
\end{equation}
where $C(n,r)$ is the number of possible combinations of $r$ objects from a set of $n$ objects.
$p_{a}^{(k)}$ is then the probability that attack $a$ is not detected by the defender.
%, and let $\widetilde{M}_a^{(k)}=0$ with probability $1-p_{a}^{(k)}$. 

\item \emph{Attack generation}.
  The adversary produces attacks by executing actions according to its policy $\{\alpha_{-1,a}^{(k)}\}_{a\in A} = \bm{\pi}_{-1}(\bm{O}_{-1}^{(k)})$ given the fully observed ADE state $\bm{O}_{-1}^{(k)}$.
  Then $M_a^{(k+1)} = \alpha_{-1,a}^{(k)}$ for each $a \in A$.
  %which is the sequence of probability of generating each type of attacks. 
%$\{\alpha_{-1,a}^{(k)}\}_{a\in A}$ is mapped from its observation $\bm{O}_{-1}^{(k)}$ of the system state by using the policy $\bm{\pi}_{-1}$.   
%For each attack type $a \in A$, let $M_a^{(k+1)}=1$ with probability $\alpha_{-1,a}^{(k)}$ and $M_a^{(k+1)}=0$ with probability $1-\alpha_{-1,a}^{(k)}$.

\item \emph{Triggering alerts}. 
Each attack $a \in A$ can trigger alerts as follows. 
For each attack $a \in A$ and alert type $t \in T$, if $M_a^{(k+1)}=1$, then $S_{a,t}^{(k+1)}=n$ with
%\Aron{Have we introduced $P$? I think that we need to say a few words about it somewhere (e.g., what it is, how it is estimated).}
probability $P_{a,t}(n)$ for $n \ge 0$.
This probability can be estimated, for example, by feeding inputs which include representative attacks into an attack detector and observing relative frequencies of alerts that are triggered.
%where $P_{a,t}(n)$ is the probability that an attack $a\in A$ triggered $n$ alerts of type $t \in T$.
%Otherwise, $S_{a,t}^{(k+1)}=0$.
In addition, false alerts are generated according to the distribution
% \Aron{Do we ever use symbol  $\mathcal{F}$? If not, then we should not introduce it (and remove from table of symbols). If yes, then we need to say a few words about it.}
$\mathcal{F}_t$, which we can estimate from data of normal behavior and associated alert counts.
%the number of which follow a probability distribution $\mathcal{F}_t$ where $\mathcal{F}_t(n)$ is the probability that $n$ false alerts of type $t$ are raised.
Let $f_t^{(k)}$ be the number of false alerts of type $t\in T$ that have been generated.
Then the total number of
%\Aron{We don't need to say ``uninvestigated'' since these alerts are all brand new.}
alerts in the next time period $k+1$ is $N_t^{(k+1)}=f_t^{(k)}+S_{a,t}^{(k+1)}$.

\end{enumerate}

In order to define the reward received by the defender in time period $k$, we make the following assumption:
\emph{if any of the alerts raised by an attack is chosen to be inspected, then the attack is detected; otherwise, the attack is not detected.}
Let~$L_a$ be the loss incurred by the defender when an attack $a\in A$ is not detected. Then the reward of the defender obtained in time period $k$ is 
\begin{equation}
R_{+1}^{(k)} = -\sum_{a \in A} L_a \cdot \widetilde{M}_a^{(k)}.
\end{equation}
For an arbitrary pure strategy profile of the defender and adversary,  $(\bm{\pi}_{+1}, \bm{\pi}_{-1})$, the defender's utility from the game is the expected total discounted sum of the reward accrued in each time period:
\begin{equation}
U_{+1}(\bm{\pi}_{+1}, \bm{\pi}_{-1}) = \mathbb{E} \left[ \sum_{k = 0}^\infty \tau^k \cdot R_{+1}^{(k)} \right],
\label{eq:utility_mdp}
\end{equation}
where $\tau \in (0, 1)$ is a temporal discounting factor which implies that future rewards are less important than current rewards.
That is, imminent losses are more important to the defender than potential future losses.
%which indicates that utilities far in the future worth exponentially less than utilities incurred at the first time slot. 
The adversary's utility is then $U_{-1}(\bm{\pi}_{+1}, \bm{\pi}_{-1}) = -U_{+1}(\bm{\pi}_{+1}, \bm{\pi}_{-1})$.
%respectively.

%\subsection{Solution Concept}
%We use \emph{mixed-strategy Nash Equilibrium (MSNE)} as the solution concept of the proposed game.

Our goal of finding robust alert investigation policies amounts to computing a \emph{mixed-strategy Nash equilibrium (MSNE)} of our game by the well-known equivalence between MSNE, maximin, and minimax solutions in zero-sum games~\cite{Korzhyk11}.
A mixed-strategy profile $(\bm{\sigma}_{v}^*,\bm{\sigma}_{-v}^*)$ of the two players is an MSNE if it satisfies the following condition for all $v \in \{+1, -1\}$
%\iAron{$ U_v(\bm{\sigma}_{v}^*,\bm{\sigma}_{-v}^*) \geq U_v(\bm{\sigma}_{v},\bm{\sigma}_{-v}^*) ~~ \forall \bm{\sigma}_{v} \in \bm{\Sigma}_v $}
\begin{equation}
U_v(\bm{\sigma}_{v}^*,\bm{\sigma}_{-v}^*) \geq U_v(\bm{\sigma}_{v},\bm{\sigma}_{-v}^*) ~~ \forall \bm{\sigma}_{v} \in \bm{\Sigma}_v. 
\label{eq:msne}
\end{equation} 
That is, each player $v$ chooses a stochastic policy $\bm{\sigma}_v^*$ that is the \emph{best response} (is optimal for $v$) when its opponent chooses~$\bm{\sigma}_{-v}^*$.
%In other words, no player can increase its utility by a unilateral deviation from its mixed strategy at MSNE.

%% file: content/game_solver.tex
\section{Computing Robust Alert\\Prioritization Policies}
\label{sec:game_solver}

\begin{figure*}[h!]
\centering
\includegraphics[width=0.65\textwidth]{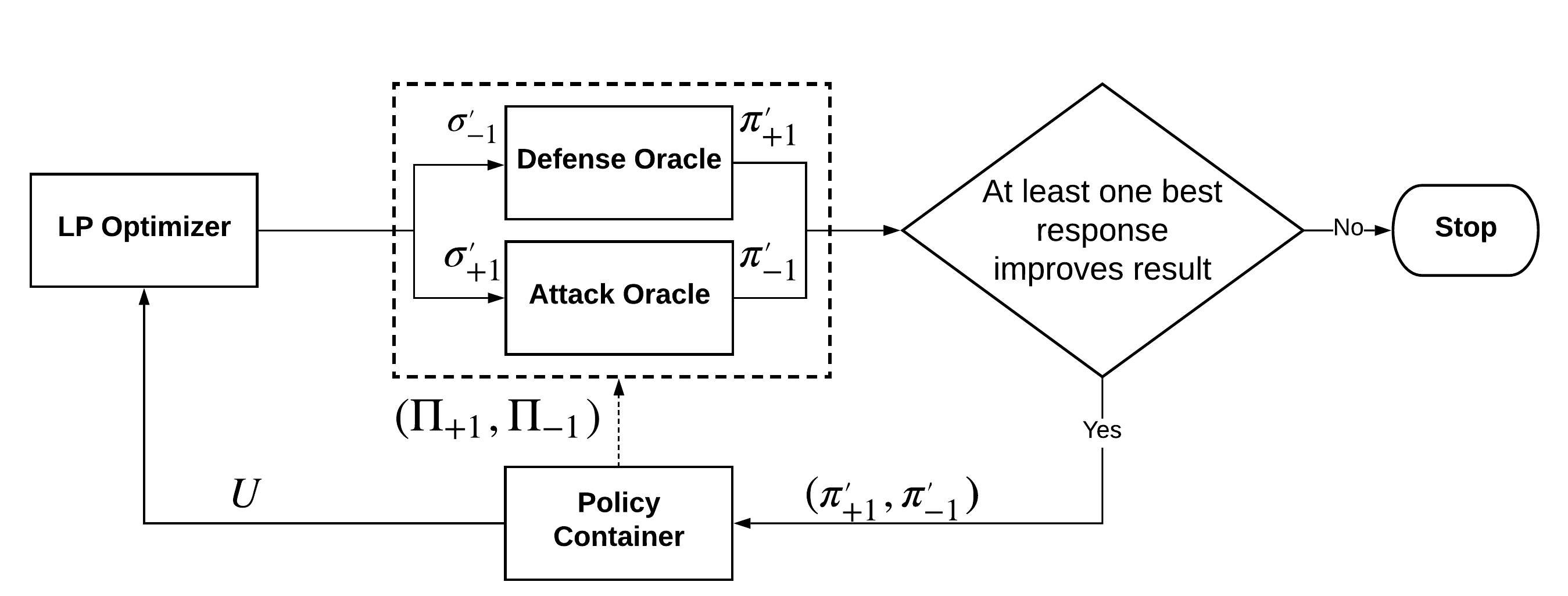}
\caption{The game solver based on the double oracle algorithm.}
\label{fig:double_oracle}
\end{figure*}

\subsection{Solution Overview}

For given sets of policies, $\bm{\Pi}_{+1}$ and $\bm{\Pi}_{-1}$, a standard approach to computing the MSNE of a zero-sum game is to solve a linear program of the following form:
%the corresponding pair of \emph{Linear Programming (LP)} problems.
%Specifically, let $U_v^*$ be the utility of player $v\in\{+1,-1\}$ at MSNE, its mixed strategy at MSNE, $\bm{\sigma}_{v}^*$, can be obtained by solving the following LP:
\begin{equation}
\begin{array}{ll@{}ll}
\text{max}  & U_v^* &\\
\text{s.t.} & \sum_{\bm{\pi}_v \in \bm{\Pi}_v} U_v(\bm{\pi}_v, \bm{\pi}_{-v})\cdot \sigma_v(\bm{\pi}_v) \geq  U_v^*, ~ & \forall \bm{\pi}_{-v} \in \bm{\Pi}_{-v} \\
                 & \sum_{\bm{\pi}_v \in \bm{\Pi}_v} \sigma_v(\bm{\pi}_v) =1                                               & \\
& \displaystyle \sigma_v(\bm{\pi}_v) \geq 0  & \forall \bm{\pi}_v \in \bm{\Pi}_v
\end{array}
\label{eq:compute_msne}
\end{equation} 
where in our case the optimal solution $\bm{\sigma}_{+1}^*$ yields the robust alert prioritization policy for the defender.
However, using this approach for our problem entails two principal technical challenges: 1) the space of policies for both players is intractably large, and 2) it is even intractable to explicitly represent individual policies, since they map a combinatorial set of states to a combinatorial set of actions for both players.

We propose an adversarial reinforcement learning approach to address these challenges, which combines a \emph{double oracle} framework~\cite{Mcmahan03} with neural reinforcement learning.
The general double oracle approach is illustrated in Figure~\ref{fig:double_oracle}.  
We start with an arbitrary small collection of policies for both players, $(\bm{\Pi}_{+1}, \bm{\Pi}_{-1})$, and solve the linear program~\eqref{eq:compute_msne}, obtaining provisional equilibrium mixed strategies $(\bm{\sigma}_{+1}, \bm{\sigma}_{-1})$ of the restricted game.
Next, we query the attack oracle to compute the adversary's best response $\bm{\pi}_{-1}(\bm{\sigma}_{+1})$ to the defender's equilibrium mixed strategy $\bm{\sigma}_{+1}$, and, similarly, query the defense oracle to compute the defender's best response $\bm{\pi}_{+1}(\bm{\sigma}_{-1})$ to the adversary's equilibrium mixed strategy $\bm{\sigma}_{-1}$.
The best response policies are then added to the policy sets $(\bm{\Pi}_{+1}, \bm{\Pi}_{-1})$ of the players, and we then re-solve the linear program and repeat the process.
The process stops when neither player's best response policy yields appreciable improvement in utility compared to the provisional equilibrium mixed strategy.
Since the space of possible policies in our case is infinite, this process may not converge.
%\Aron{Ambiguos! This might be interpreted as suggesting that our policy space is finite (which is not true). Better write: ``If the policy space if finite, ...''}Given that the policy space is finite, the process is guaranteed to converge in finite time, and in practice requires relatively few iterations.
However, in our experiments the procedure converged in fewer than 15 iterations (see Figure~\ref{fig:execution_time} in Appendix~\ref{sec:execution_cost}),
%\Aron{Maybe we should make it clear here that we don't have converge guarantee, but our experimental results are awesome.},
with the fast convergence in part due to the way we represent policies, as discussed below.
The main question that remains is how to compute or approximate the best response oracles for both players.
To this end, we use reinforcement learning techniques with policies represented using neural networks.
Below, we explain both our double oracle approach and our neural reinforcement learning methods (including the specific way in which we represent policies) in further detail.

%In practice, such a premise condition does not always hold due to the massive policy space of the two players.
%As a policy of each player is a mapping from a state into action, such mapping can be represented by infinite ways.
%Consequently, it incurs significant computational complexity if we explicitly enumerate the entire set of policy and solve the corresponding LP problem.
%To address this challenge, we extend the conventional \emph{Double Oracle algorithm} \cite{tsai2012} in game theory to policy space so as to approximate the mixed-strategy Nash Equilibrium.
%The basic idea is that we start with a restricted zero-sum game with the subsets of the policies for the players, then iteratively increase the set of policies by computing the %mixed-strategy Nash Equilibrium of the restricted game and adding the pure-strategy best response of the equilibrium into the set of policies.
%The MSNE obtained at the equilibrium is applied as an approximate of the MSNE of the game.
%Next, we describe our solution in details. 

\subsection{Policy-based Double Oracle Method}
As displayed in Figure~\ref{fig:double_oracle}, our game solver is an extension of the double oracle algorithm proposed in \cite{tsai2012} and is partitioned into four parts: a policy container, a linear programming (LP) optimizer, a defense oracle, and an attack oracle.
The policy container stores the policies of the two players, $\bm{\Pi}_{+1}$ and $\bm{\Pi}_{-1}$, as well as a utility matrix $\bm{U}$, whose elements are $U_{+1}(\bm{\pi}_{+1}, \bm{\pi}_{-1})$ for all $\bm{\pi}_{+1} \in \bm{\Pi}_{+1}$ and $\bm{\pi}_{-1} \in \bm{\Pi}_{-1}$.
The LP optimizer solves the game by computing the current mixed-strategy Nash equilibrium given the utility matrix $\bm{U}$.
The defense and attack oracles are agents that apply reinforcement learning to compute the optimal responses to their opponents' mixed strategies, which are provided by the LP optimizer.

Our solver works in an iterative manner such that the players' policies and the utility matrix grow incrementally.
Initially, $\bm{\Pi}_{+1}$, $\bm{\Pi}_{-1}$ can be set up with some basic policies, for example, uniformly allocating each player's budget among their choices.
Then, the policy sets,  jointly encapsulated in a \emph{policy container}, are updated in each iteration as follows:
\begin{enumerate}

%\item \Aron{The description in this subsection is way to detailed. We have already described double oracle. We could, for example, remove this first point entirely.} First, the policy container forwards the players' sets of policies, $( \bm{\Pi}_{+1}, \bm{\Pi}_{-1})$, to the defense oracle and attack oracle as well as the LP optimizer, and also forwards the corresponding utility matrix $\bm{U}$ to the LP optimizer.

\item First, the LP optimizer computes the mixed-strategy Nash Equilibrium $(\bm{\sigma}_{+1}^{'}, \bm{\sigma}_{-1}^{'})$ of the current iteration by solving the optimization problems presented in Equation (\ref{eq:compute_msne}).

\item The oracle of player $v$ computes the best response policy~$\bm{\pi}'_{v}$ given that its opponent uses its equilibrium mixed-strategy $\bm{\sigma}'_{-v}$, for $v \in \{+1, -1\}$. 

\item If $U_v(\bm{\pi}'_{v}, \bm{\sigma}'_{-v}) \leq U_v(\bm{\sigma}'_{v},\bm{\sigma}'_{-v})$ for all $v \in \{+1,-1\}$, the double oracle algorithm terminates and returns $(\bm{\sigma}'_{+1}, \bm{\sigma}'_{-1})$ as the approximate MSNE. Otherwise, add~$\bm{\pi}'_v$ to the corresponding $\bm{\Pi}_v$, update the utility matrix $\bm{U}$ and continue from Step 2.
%In other words, the double oracle algorithm terminates when neither of the best response policies is superior to the given player’s current mixed strategy against the fixed opponent mixed strategy at current MSNE.
\end{enumerate}
The resulting $(\bm{\sigma}_{+1}^{'}, \bm{\sigma}_{-1}^{'})$ is an approximate mixed-strategy Nash equilibrium $(\bm{\sigma}_{+1}^{*}, \bm{\sigma}_{-1}^{*})$.
%It has shown that $(\bm{\sigma}_{+1}^{'}, \bm{\sigma}_{-1}^{'})$ can admit a quality guarantee such that
%\begin{equation}
%U_{+1}(\bm{\sigma}_{+1}^{'}, \bm{\sigma}_{-1}^{'}) \geq \beta\cdot U_{+1}(\bm{\sigma}_{+1}^{*}, \bm{\sigma}_{-1}^{*})
%\end{equation}  
%for $\beta\in (0,1)$~\cite{tsai2012}.

Next, we describe how the defense and attack oracles apply neural reinforcement learning to compute their best responses to an arbitrary mixed-strategy of the opponent.  

\subsection{Approximate Best Response Oracles with Neural Reinforcement Learning}

We now turn to our approach to compute $\bm{\pi}'_{v}$, the optimal response of player $v$ when its opponent uses a mixed strategy~$\bm{\sigma}'_{-v}$ such that
\begin{equation}
\bm{\pi}'_v = \argmax_{\bm{\pi}_v} U_v(\bm{\pi}_v, \bm{\sigma}'_{-v}).
\label{eq:defender_br}
\end{equation}
%where
%\begin{equation}
%\begin{split}
%U_v(\bm{\pi}_v, \bm{\sigma}'_{-v}) &= \sum_{\bm{\pi}_{-v}\in\bm{\Pi}_{-v}}\sigma'_{-v}(\bm{\pi}_{-v}) U_v(\bm{\pi}_v, \bm{\pi}_{-v}) \\
%&= \sum_{\bm{\pi}_{-v}\in\bm{\Pi}_{-v}}\sigma'_{-v}(\bm{\pi}_{-v}) \mathbb{E} \left[ \sum_{k = 0}^\infty \tau^k \cdot R_{+1}^{(k)} \right]
%\end{split}
%\end{equation}
This problem poses a major technical challenge, since the spaces of possible policies for both the defender and the attacker are quite large.
To address this, we propose using the reinforcement learning (RL) paradigm.
%The above problem can be solved with reinforcement learning, a paradigm widely applied to solve decision problems when an agent takes actions in an environment to maximize its utility.
However, the use of RL poses two further challenges in our setting.
First, for a given state, each player's set of possible actions is combinatorial.
For example, the attacker is choosing subsets of attacks, whereas the defender is choosing subsets of alerts.
Consequently, we cannot use common methods such as \emph{Q-learning}, which requires explicitly representing the action-value function $Q(x,a)$ for every possible action $a$, even if we approximate this function over states $x$ using, e.g., a neural network, as is common in deep RL.
We can address this issue by appealing to \emph{actor-critic} methods for RL, where the policy is represented as a parametric function $\bm{\pi}_{v;\theta}$ with parameters $\theta$. However, this brings up the second challenge: actor-critic approaches learn policies using gradient-based methods, which require that the actions are continuous.
In our case, however, the actions are discrete.

One solution is to learn the action-value function $Q(x,a)$ over a vector-based representation of actions, such as using a binary vector to indicate which attacks are used.
The problem with this approach, however, is that the resulting policy $\bm{\pi}_{v} \in \arg\max_{a \in A} Q(x,a)$ is hard to compute in real time, since it involves a combinatorial optimization problem in its own right.
We therefore opt for a much more scalable solution that uses the actor-critic paradigm with an alternative representation of the adversary and defender policies, which admits gradient-based learning.

%\Aron{Kinda informal, conversational. ``First, consider the adversary.''}
Let us start with the adversary.
Recall that the adversary's policy maps a state to a subset of attack actions
%\Aron{We have already switched to the vector representation $\bm{\alpha}$? Why are we considering the set representation $\tilde{A}$ again?}
$A$, with the constraint on the total budget used by the chosen actions.
Instead of returning a discrete subset of actions, we map the adversary's policy to a \emph{probability distribution} over actions, overloading our prior notation so that $\alpha_{-1,a}^{(k)}$ now denotes the \emph{probability} that action $a \in A$ is executed.
Now the policy can be used with actor-critic methods, but it may violate the budget constraint.
To address this final issue, we simply project the probability distribution into the feasible space at execution time by normalizing it by the total cost of the distribution, and then multiplying by the budget constraint.
Notice that in this process we have relaxed the attacker's budget constraint to hold only in \emph{expectation}; however, this only makes the attacker stronger.
An interesting side-effect of our transformation of the adversary's policy space is that the RL method will now effectively search in the space of \emph{stochastic} adversary policies.
An associated benefit is that it leads to faster convergence of the double oracle approach.

Next, consider the defender.
In this case, we can simply represent the policy as a mapping to fractions of the \emph{total defense budget} allocated to each alert type $t$.
In other words, for each alert type $t$, the policy will output the maximum fraction of the defense budget that will be used to inspect alerts of type $t$.
%\Aron{I'm not sure if this will be clear to the reader.}
This simultaneously makes the mapping continuous, and obviates the need to explicitly deal with the budget constraint.

The final nuance is that RL methods are typically designed for a fixed environment, whereas our setting is a game.
However, note that since we are concerned only with each player's best response to the other's mixed strategy, we can embed the mixed strategy of the opponent as a part of the environment.
Next, we describe our application of actor-critic methods to our problem, given the alternative representations of adversary and defender policies above.

%First, the action space of each player is continuous thus traditional \emph{value-based} approaches in reinforcement, such as \emph{Q-learning}, cannot be applied directly.\footnote{The action of the defender is a sequence of integers, which are treated as continuous in our paper by presenting them as real-valued fractions of the defender's budget and then convert them back to integers.}
%These methods typically learn a \emph{state-value function} which represents the value of choosing an action in a state.
%Then, they obtain a corresponding policy by greedily choosing the action that maximizes the state-value function.
%When the action space is continuous,  these paradigms becomes intractable as it is infeasible to enumerate all the actions and greedily choose the best one.
%Second, the existence of the opponent with a stochastic policy is not considered by conventional reinforcement learning approaches, leaving these methods intractable in our settings.

\begin{figure}[t]
\centering
\includegraphics[width=0.4\textwidth]{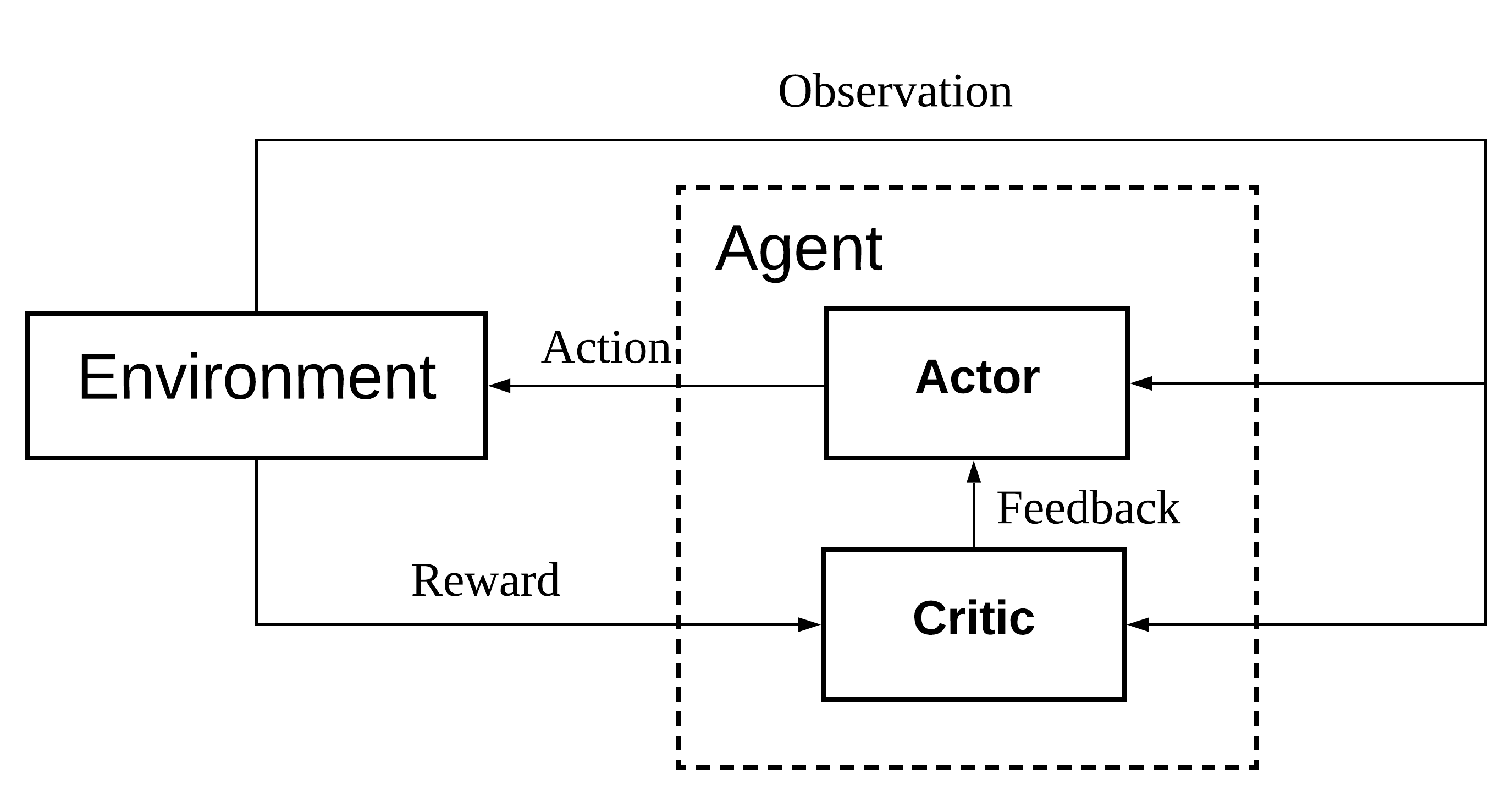}
\caption{The interactions among actor, critic and environment.}
\label{fig:actor_critic}
\end{figure}

%To address the first challenge, we applied the \emph{actor-critic} paradigm in reinforcement learning.
The basic idea of the actor-critic method is that we can iteratively learn and improve a policy without enumerating actions by using two parallel processes that interact with each other:  an actor which develops a policy, and a critic network which evaluates the policy.
The interaction between the actor and critic in illustrated in Figure~\ref{fig:actor_critic}. 
In each iteration, the actor and critic proceed as follows: 
\begin{enumerate}
\item The actor executes an action according to its policy given the observation of the environment.
\item Upon receiving the action, the environment updates its system state and returns a reward to the critic.
\item The critic updates its evaluation method and provides feedback to the actor.
\item The actor updates its policy according to the feedback given by the critic.
\end{enumerate}
%To address the second challenge, we extend the conventional actor-critic approach to multiagent settings.

We propose \emph{DDPG-MIX}, actor-critic algorithm that operates in continuous action spaces and computes an approximate best response to an opponent who uses a stochastic policy.
DDPG-MIX is an extension of the \emph{Deep Deterministic Policy Gradient (DDPG)} approach proposed in \cite{lillicrap2016} to our setting, and the full algorithm is outlined in Algorithm \ref{alg:ddpg_mix} in the Appendix.
%\Aron{``in Appendix ?'' (because it's nowhere near)}.
For each player~$v$, DDPG-MIX employs two neural networks to represent the actor and critic: a policy network $\bm{\pi}_{v}(\bm{O}_v|\bm{\theta}_v^\pi)$ for the actor, which has parameters $\bm{\theta}_v^\pi$ and maps an observation $\bm{O}_v$ into an action, and a value network $Q_{v}(\bm{O}_v, \bm{\alpha}_v|\bm{\theta}_v^Q)$ for the critic, which has parameters $\bm{\theta}_v^Q$ and maps an observation $\bm{O}_v$ and an action $\bm{\alpha}_v$ into a value.
Initially, these two neural networks are randomly initialized.
Then, we train these two iteratively with multiple episodes, each of which contains multiple steps.
At the beginning of each episode, the opponent samples a deterministic policy $\bm{\pi}_{-v}$ with its mixed-strategy $\bm{\sigma}_{-v}$.
The policy network and value network are then updated as follows.
First, we generate an action by using the $\epsilon$-greedy method: we randomly choose an action with probability $\epsilon$ (called  \emph{exploration} in RL), and apply the policy network $\bm{\pi}_{v}(\bm{O}_v|\bm{\theta}_v^\pi)$ to produce an action corresponding to the current state with probability $1 - \epsilon$ (called \emph{exploitation}).
Player~$v$ then executes the action produced and so does its opponent, which executes an action $\bm{\alpha}_{-v}$ returned by $\bm{\pi}_{-v}$.
Once the system state of the environment is updated,  player~$v$ receives the reward and stores the transition into a memory buffer. 
Player~$v$ then samples a %\Aron{Buffer, minibatch, etc. are not explained. Will the reader be able to follow this?} 
minibatch, a subset of transitions randomly sampled from the buffer, to update the value network $Q_{v}(\bm{O}_v, \bm{\alpha}_v|\bm{\theta}_v^Q)$ by minimizing a loss function as in most regression tasks.
The sampled gradient of the value network with respect to $\bm{\alpha}_v$ is then forwarded to the policy network, which is further applied to update $\bm{\pi}_{v}(\bm{O}_v|\bm{\theta}_v^\pi)$ as presented in Equation \eqref{eq:ddpg_grad} in Algorithm \ref{alg:ddpg_mix}. 
After a fixed number of episodes, the resulting policy network $\bm{\pi}_{v}(\bm{O}_v|\bm{\theta}_v^\pi)$ is returned as the parameterized optimal response to an opponent with mixed-strategy $\bm{\sigma}_{-v}$.

\subsection{Preprocessing}
\label{S:preprocessing}

An important consideration in applying the above approaches is scalability of training.
One way to significantly improve scalability is through preprocessing, and pruning alerts for which the (near-)optimal decision is obvious.
We use the following pruning step to this end.
Suppose that there is an alert type $t$ which is generated by benign traffic with probability at most $\epsilon$, where $\epsilon$ is very small (for example, $\epsilon=0$, in which case alerts of type $t$ \emph{never} correspond to a false positive).
In most realistic cases, it is nearly optimal to always inspect such alerts.
Consequently, we prune all alerts with false positive rate below a small pre-defined $\epsilon$ (in our implementation below, we set $\epsilon = 0$), and mark them for inspection (correspondingly reducing the available budget for inspecting other alerts).

%% file: content/case_studies.tex
\section{Case studies}
\label{sec:case_studies}

In this section, we present case studies to investigate the robustness of our proposed approach for alert prioritization.
We conduct our experiments in two applications: intrusion detection
which employs a signature-based detection system and fraud detection which applies a learning-based detection system.
We start with a broad introduction of the experimental methodology,
including the details of the implementation of our approach and
evaluation methods.
%including evaluation approach and implementation of our method.
%including descriptions of the datasets, attack detectors, baseline alternatives, evaluation metrics, and our implementations of the proposed approach.
We then proceed to describe each case study in detail.

\subsection{Experimental Methodology}
\input{content/experimental_methodology}

\subsection{Case Study I: Intrusion Detection}
\input{content/case_study_ids}

\subsection{Case Study II: Fraud Detection}
\input{content/case_study_fraud}

%% file: content/experimental_methodology.tex
%\subsubsection{Datasets}

%We conducted our experiments using two datasets: the fraud dataset and the CICIDS2017 dataset.
%We now describe these in detail.

%\subsubsection{Attack Detectors}
%We use the following two detectors in our experiments.

%\subsubsection{Baseline Approaches}
%\label{sec:baseline_approaches}

\subsubsection{Implementation}

\begin{table*}[t]
\caption{Architecture of the implemented policy and value networks.}
\centering
\scalebox{0.85}{
\begin{tabular}{|c|c|c|c|c|}
\hline
\textbf{Neural  network}         & \textbf{Layer} & \textbf{Number of units}                                   & \textbf{Activation function} & \textbf{Initializer} \\ \hline\hline
\multirow{3}{*}{Policy  network} & Input          & $T$ (defender); $|T|+|A|\cdot(1+|T|)$ (adversary)         & -                            & -                    \\ \cline{2-5} 
                                 & Hidden         & 16 (Fraud detection);  32 (Intrusion detection)                             & Tanh                         & Xavier~\cite{glorot2010xavier}               \\ \cline{2-5} 
                                 & Output         & $|T|$ (defender); $|A|$ (adversary)                       & Sigmoid                      & Xavier               \\ \hline
\multirow{3}{*}{Value  netwrok}  & Input          & $2\cdot|T|$ (defender); $|T|+|A|\cdot(2+|T|)$ (adversary) & -                            & -                    \\ \cline{2-5} 
                                 & Hidden         & 32 (Fraud detection);  64 (Intrusion detection)                             & Relu                         & He Normal~\cite{he2015initializer}            \\ \cline{2-5} 
                                 & Output         & 1                                                          & Relu                         & He Normal            \\ \hline
\end{tabular}
}
\label{tab:nn}
\end{table*}

The DDPG-MIX algorithm was implemented in TensorFlow~\cite{abadi2016tensorflow}, an open-source library for neural network learning.
The architecture of the policy and value networks for both players are displayed in Table~\ref{tab:nn}.
We used Adam for learning the parameters of the neural networks with learning rates of 0.001 and 0.002 for the policy and value networks, respectively.
The discount factor $\tau$ was set to be 0.95, and we set the size of the memory buffer to 40,000.
The learning process contained 500 episodes, each with 400 learning steps.  
The collection of policies used in the double-oracle framework was initialized with a pair of policies that uniformly allocate each player's budget among their choices.%\Aron{Uniform has not been introduced at this point!}

Our experiments were conducted on a server running Ubuntu 16.04 with Intel(R) Xeon(R) CPU E5-2695 v4 @ 2.10GHz, 18 cores and 64 GB memory.
Each experiment was repeatedly executed 20 times with 20 different random seeds.

\subsubsection{Evaluation Method}
We use the expected loss of the defender (equivalently, gain of the adversary) as the metric throughout our evaluation.
%The loss is the opposite number of the defender's utility and is equivalent to the gain of the adversary.
Specifically, for a given defense policy, we evaluate the loss of the defender using several models of the adversary.
First, we used Algorithm~\ref{alg:ddpg_mix} to compute the best response of the adversary, as anticipated by our approach.
In addition, to evaluate the general robustness of our approach, we employed two alternative policies for the adversary:
\emph{Uniform}, a policy which uniformly distributes the adversary's budget over attack actions; 
and \emph{Greedy}, a policy which allocates the budget to attacks in the order of expected adversary utility.
Specifically, the \emph{Greedy} adversary prioritizes the attack actions according to $L_a\cdot \min\{\frac{\tilde{D}}{c_a}, 1\}$, where $\tilde{D}$ is the available attack budget, adding actions in this priority order until the adversary's budget is exhausted.
%Then, the attacker exhausts its budget based on this prioritization. 

We first conduct our experiments by assuming that the defender knows the adversary's capabilities.
Subsequently, we evaluate the robustness of our approach when the defender is uncertain about the adversary's capabilities, and use it to provide practical guidance.
We also provide results on the computational cost of our approach in Appendix~\ref{sec:execution_cost}.
%experiment in cases when the defender is uncertain about the capabilities of the adversary.
%Specifically, when both players choose stochastic policies with reinforcement learning,  we use the loss of the defender at MSNE.
%When one of the players chooses a stochastic policy, we use the loss of defender associated with the utility obtained by using Equation (\ref{eq:utility_pure_mix}) and (\ref{eq:utility_mdp}).
%When neither of the players uses a stochastic policy, the loss of defender is obtained by computing the associated utility with Equation (\ref{eq:utility_mdp}).

%% file: content/case_study_ids.tex
% We now turn to our second case study on intrusion detection.
Our first case study involves a signature-based intrusion detection scenario, using the Suricata, a state-of-the-art open source intrusion detection system (IDS), combined with the CICIDS2017 dataset.
%We investigate the robustness of the proposed approach by prioritizing alerts produced by Suricata.
Our case study evaluates our alert prioritization method in two cases: i) the defender has full knowledge of the adversary; and ii) the defender is uncertain about the adversary's capabilities.

\subsubsection{CICIDS2017 dataset}

The CICIDS2017 dataset~\cite{sharafaldin2018}  records benign and malicious network flows in pcap format, captured in a real-world network between 07/03/2017 and 07/27/2017.
The network consists of 10 desktops belonging to regular users and 5 laptops owned by attackers.
The desktops are used to generate natural benign background traffic by using a profile system that abstracts the behaviors of regular users.
The laptops are employed to produce malicious traffic of the following classes of attacks: Brute Force, Botnet, DDoS, DoS, Heartbleed, Infiltration, Portscan, and Web Attack.

\subsubsection{Suricata IDS}

We employ Suricata\footnote{Available at https://suricata-ids.org/about/open-source/.} to conduct our case study on the CICIDS2017 dataset.
Suricata is an open-source network intrusion detection system which performs analysis of passing traffic on a network by using a set of signatures (also called rules).
If a traffic pattern matches any of the signatures, then a corresponding alert is triggered and sent to the network administrator.

A Suricata signature contains the following parts: \emph{action}, \emph{header}, \emph{rule options}, and \emph{priority}.
\emph{Action} describes the operation of Suricata when a signature is matched, which can be either dropping a packet or raising an alert.
\emph{Header} defines the protocol, port, and IP addresses of the source and destination in a signature.
\emph{Rule options} include a list of keywords, for example, the corresponding alert type associated with a priority.  
Finally, the \emph{priority} keyword comes with a numerical value ranging from 1 to 255 where 1 indicates the highest priority and 255 the lowest.

In our experiments, we use Suricata to scan the pcap files in the CICIDS2017 dataset.
Specifically, we use the \emph{Emerging Threats Ruleset (ETR)}\footnote{Available at \url{https://rules.emergingthreats.net/open/suricata/}.} to analyze the network traffic in the dataset.
ETR defines a total of 33 alert types, and we select the 10 most common alert types exhibited during our experiments, which are shown in Table \ref{tab:suricata_alert}.

\begin{table}[t]
\caption{Alert types of Suricata in our experiments.}
\centering
\scalebox{0.85}{
\begin{tabular}{|l|l|l|}
\hline
\textbf{Alert type}              & \textbf{Description}                           & \textbf{Priority} \\ \hline\hline
attempted-recon         & Attempted Information Leak            & 2        \\ \hline
attempted-user          & Attempted User Privilege Gain         & 1        \\ \hline
bad-unknow              & Potentially Bad Traffic               & 2        \\ \hline
misc-acticity           & Misc activity                         & 3        \\ \hline
not-suspicious          & Not Suspicious Traffic                & 3        \\ \hline
policy-violation        & Potential Corporate Privacy Violation & 1        \\ \hline
protocol-command-decode & Generic Protocol Command Decode       & 3        \\ \hline
trojan-activity         & A Network Trojan was Detected         & 1        \\ \hline
unsuccessful-user       & Unsuccessful User Privilege Gain      & 1        \\ \hline
web-application-attack  & Web Application Attack                & 1        \\ \hline
\end{tabular}
}
\label{tab:suricata_alert}
\end{table}

\subsubsection{Experimental Setup}

\begin{table*}[t]
\centering
\caption{Attack actions and alert types used in the case study of intrusion detection.}
\scalebox{0.9}{
\begin{tabular}{|c|c|c|c|c|c|c|c|c|c|}
\hline
\multirow{2}{*}{\textbf{Attack action}} & \multicolumn{7}{c|}{\textbf{Number of each alert type raised}}                                                               & \multirow{2}{*}{$E_a$} & \multirow{2}{*}{$L_a$} \\ \cline{2-8}
                                        & attempted-recon & attempted-user & bad-unknown & misc-activity & not-suspicious & policy-violation & protocol-command-decode &                        &                        \\ \hline
Brute Force                             & 1230            & 0              & 0           & 0             & 0              & 0                & 0                       & 120                    & 3.6                    \\ \hline
Botnet                                  & 0               & 4              & 2           & 106           & 0              & 54               & 0                       & 60                     & 6.0                    \\ \hline
DoS                                     & 0               & 0              & 0           & 0             & 0              & 24               & 0                       & 74                     & 4.0                    \\ \hline
Heartbleed                              & 0               & 0              & 4           & 0             & 10             & 0                & 0                       & 20                     & 3.6                    \\ \hline
Infiltration                            & 710             & 2              & 862         & 12            & 0              & 80               & 600                     & 52                     & 1.4                    \\ \hline
PortScan                                & 138             & 0              & 320         & 30            & 0              & 0                & 0                       & 80                     & 1.4                    \\ \hline
Web Attack                              & 0               & 0              & 6           & 0             & 0              & 0                & 0                       & 62                     & 2.7                    \\ \hline
\end{tabular}
}
\label{tab:ids_attack_alert}
\end{table*}

\begin{table}[t]
\centering
\caption{Average number of false alerts triggered in each time period}
\scalebox{0.9}{
\begin{tabular}{|c|c|}
\hline
\textbf{Alert type}     & \textbf{Avg. number of false alerts in each period} \\ \hline
attempted-recon         & 7,200                                                \\ \hline
attempted-user          & 44,100                                               \\ \hline
bad-unknown             & 1,600                                                \\ \hline
misc-activity           & 7,300                                                \\ \hline
not-suspicious          & 17,400                                               \\ \hline
policy-violation        & 4,000                                                \\ \hline
protocol-command-decode & 10,200                                               \\ \hline
\end{tabular}
}
\label{tab:ids_false_alert}
\end{table}

We use the following steps to set up our experiments for the case study.
First, we used 30 minutes as the fixed length of each time period.
Then, we utilized the Suricata IDS to scan and detect intrusions for both malicious and benign traffic in the CICIDS2017 data.
By doing so, we obtained the number of alerts of each type raised by each attack action,
%different types of attack actions,
%\footnote{The alerts produced with Suricata are different from those raised in the fraud detector: In fraud detection, each attack action can raise at most one alert for each type. However, a single attack action in intrusion detection can raise multiple alerts each of which has the same type. (e.g. a single Infiltration attack may trigger hundreds of the attempted-recon alerts in Suricata.)}
as well as the number of false alerts in each time period.
%Similar to our case study on fraud detection,
In the preprocessing step we pruned alert types that were triggered \emph{only} by malicious traffic, as discussed in Section~\ref{S:preprocessing}.
%, since these can be trivially used to detect attacks without our complex machinery (we can simply implement this as a preprocessing step before using our alert prioritization policy).
As a result, we were left with 7 out of the 10 alert types to consider using our full adversarial RL framework.
In addition, we filtered out the attack actions that do not raise any alerts, since those attacks will never be detected using Suricata, leaving 7 out of 8 representative attacks for our experiments.
The final attack actions and alert types that we use in the experiments are given in Table~\ref{tab:ids_attack_alert}.

We used Poisson distribution to fit the distribution of alerts raised by benign traffic in each time period.
Since the benign traffic in the CICIDS2017 dataset was captured from only 10 desktop which is far less than the number of computers in a real-world local area network, we amplified the corresponding mean of each type of alert by a factor of 100.
The resulting average numbers are shown in Table~\ref{tab:ids_false_alert}.
We set the cost of investigating each alert to 1.0 (i.e., equal for all alerts).
Next, we used the base score of the \emph{Common Vulnerability Scoring System (CVSS)} to measure the loss of defender if an attack action was not detected.
Specifically, we employed CVSS v3.0\footnote{Available at https://www.first.org/cvss/calculator/3.0.} to compute $L_a$ for $a\in A$.
Note that since the defender observes only \emph{alerts} but not the actual attacks, alert-investigation decisions in deployment cannot directly take advantage of the CVSS scores to quantify the risk of underlying attack. However, since the ground truth is available during training and evaluation, CVSS scores are used to provide additional information on the impact of the attack. 
% \emph{Note that we do not assume that CVSS scores are known at decision-time}: in fact, they cannot be, since the defender only observes \emph{alerts}, and not actual attacks. Rather, we use them to evaluate relative effectiveness of alternative policies in our experiments. Moreover, we used the execution time of each attack action as the cost of mounting this attack. 
For example, the cost of mounting a Brute Force attack is 120 minutes.
%, and since the length of a time period is 30 minutes, this implies that the adversary needs 4 (out of 5) laptops to execute this attack in the allotted time.
We document $L_a$ (loss to the defender from a successful attack) and $E_a$ (execution cost of the attack) for $a\in A$ in Table~\ref{tab:ids_attack_alert}. 

\subsubsection{Baselines}

The performance of the proposed approach is compared with two alternative policies for alert prioritization:
\emph{Uniform}, a policy which uniformly allocates the defender's budget over alert types, and 
%\emph{GAIN}~\cite{laszka2017}, a game theoretic approach which prioritizes alerts and exhausts the defender's budget according to the prioritization; \emph{RIO}~\cite{yan2018}, another game theoretic approach which prioritizes alerts and determines a threshold for the maximum budget that the defender can distribute on each type of alerts;
\emph{Suricata} priorities, where the defender exhausts the defense budget according to the built-in prioritization of the Suricata IDS, shown in Table~\ref{tab:suricata_alert}.
We tried two additional baselines from prior literature that use game theoretic alert prioritization: 
\emph{GAIN}~\cite{laszka2017} and \emph{RIO}~\cite{yan2018}, but these do not scale computationally to the size of our IDS case study (we compare to these in our second, smaller, case study).
We did not compare to alert correlation methods for reducing the number of false alerts, since these techniques are entirely orthogonal and complementary to our setting (we address the issue of limited alert inspection budget in the face of false alerts, whatever means are used to generate alerts).
Throughout, we refer to our proposed approach as \emph{ARL}.

%Since GAIN and RIO were unable to scale to cases when the volume of alerts and defense budget are very large ($>$10,000), we employed Uniform and Suricata as the alternative baselines for the defender in our case study.

\subsubsection{Results}
\label{sec:case_study_ids}

\begin{figure}[h!]
\centering
\begin{tabular}{cc}
  \includegraphics[width=0.22\textwidth]{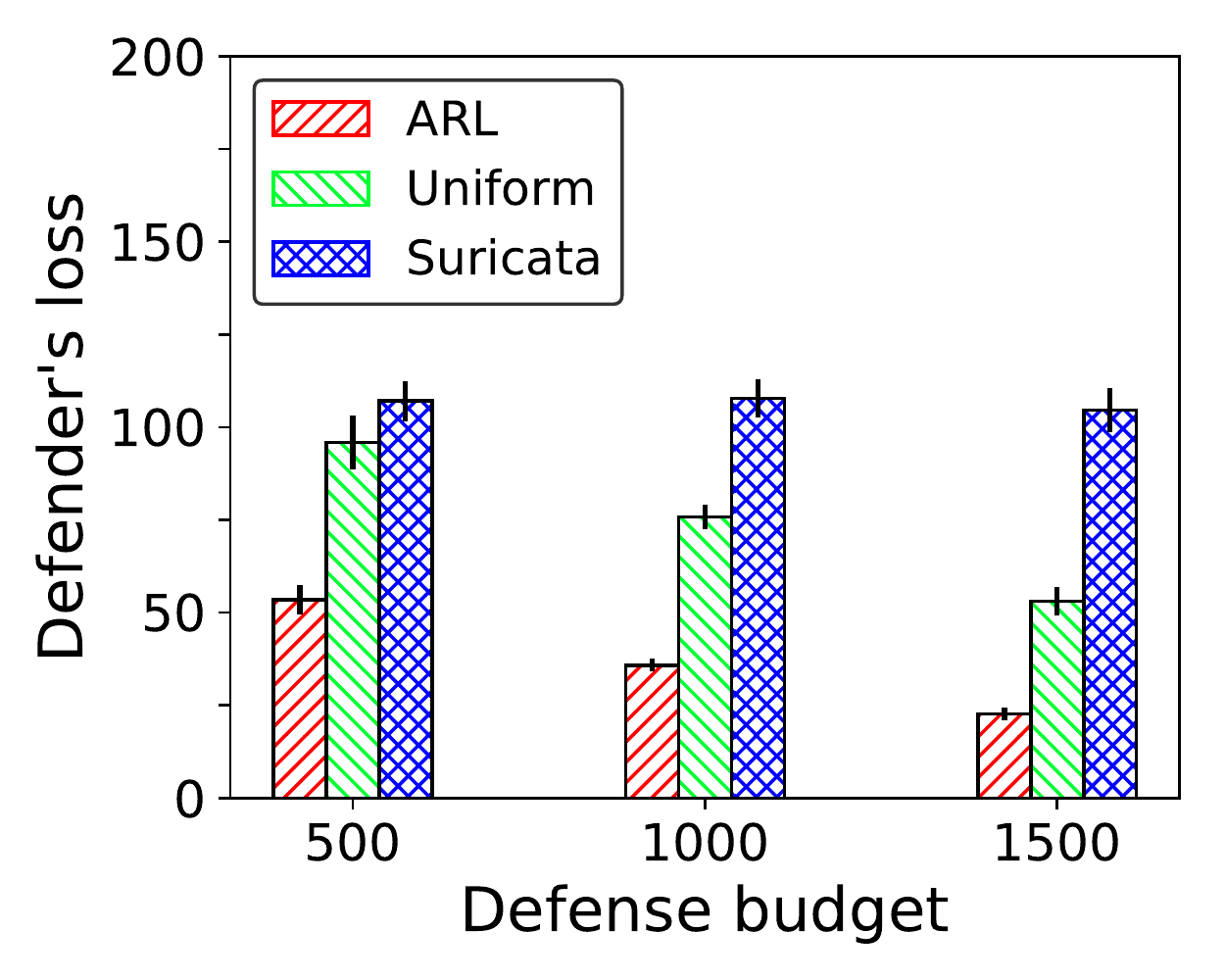} &
 \includegraphics[width=0.22\textwidth]{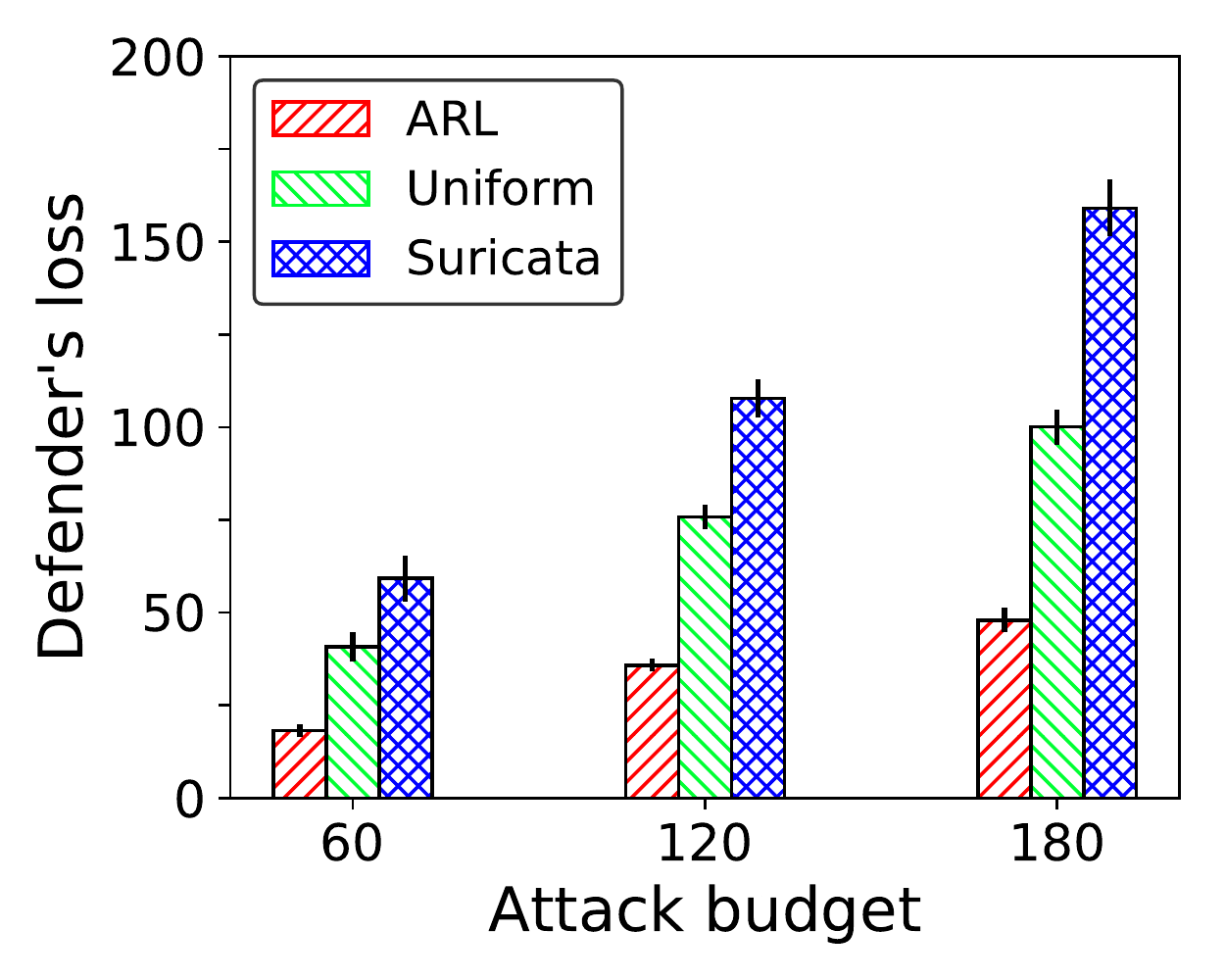}\\
\end{tabular}
	\caption{Intrusion detection: loss of the defender when it knows the attack budget.  Left: Defender's loss for different defense budgets, with attack budget fixed at 120.  Right: Defender's loss for different attack budgets, with defense budget fixed at 1000.}
	\label{fig:ids_complete}
      \end{figure}

      Figure~\ref{fig:ids_complete} presents our evaluation of the robustness of alert prioritization approaches when the defender knows the adversary's capabilities, and the results suggest that our approach significantly outperforms the other baselines.
      %, and the results are consistent with our observations for fraud detection.
      Specifically, the proposed approach is $50\%$ better than the Uniform policy, which in turn is significantly better than using Suricata priorities.
      There are a few reasons why deterministic priority-based approaches perform so poorly.
      First, determinism allows attackers to easily circumvent the policy by focusing on attacks that trigger alerts which are rarely inspected.
      Moreover, such naive deterministic policies also fail to exploit the empirical relationships between attacks and alerts they tend to trigger: for example, if an attack triggers multiple alerts, but one of these alert types happens to have very few alerts in current logs, static priority-based policies will not leverage this structure.
      In contrast, by learning a \emph{policy} of alert inspection which maps arbitrary alert observations to a decision about which to inspect, we can make decisions at a significantly finer granularity.
%In contrast, the built-in Suricata prioritization fails to provide adequate robustness: it even leads to higher loss of the defender compared to Uniform, a naive Uniform policy.
%The major reason is that Suricata exhausts the defense budget over alerts with higher priorities, which is a waste to its budget.
%To detect an attack action in the case of intrusion detection, it is unnecessary to investigate all the alerts of a specific type, even this type has the highest priority.
%Since an attack action can trigger multiple alerts with the same type, if any of these are investigated, then the attack is revealed.

\begin{figure}[h!]
\centering
\begin{tabular}{cc}
  \includegraphics[width=0.22\textwidth]{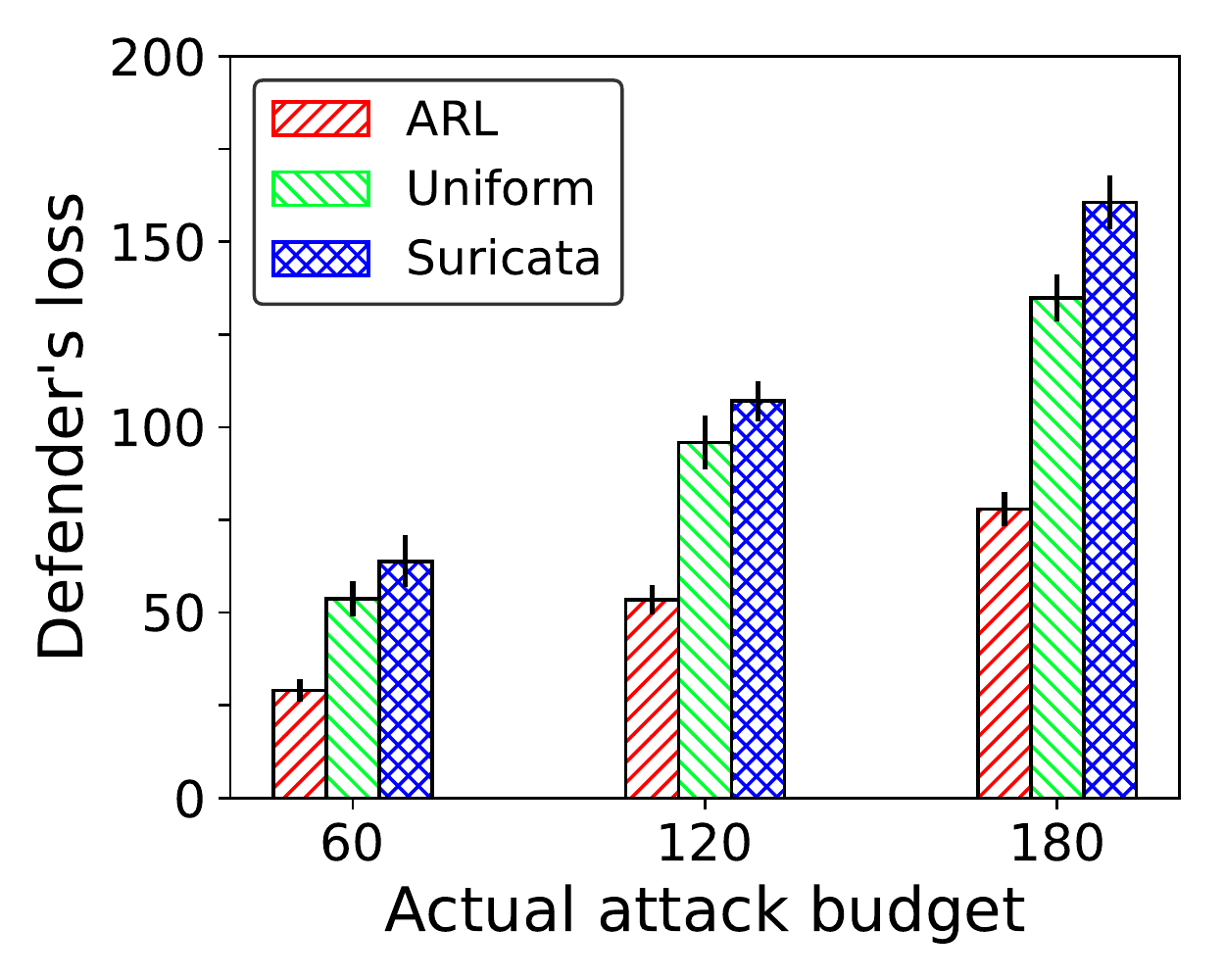} &
 \includegraphics[width=0.22\textwidth]{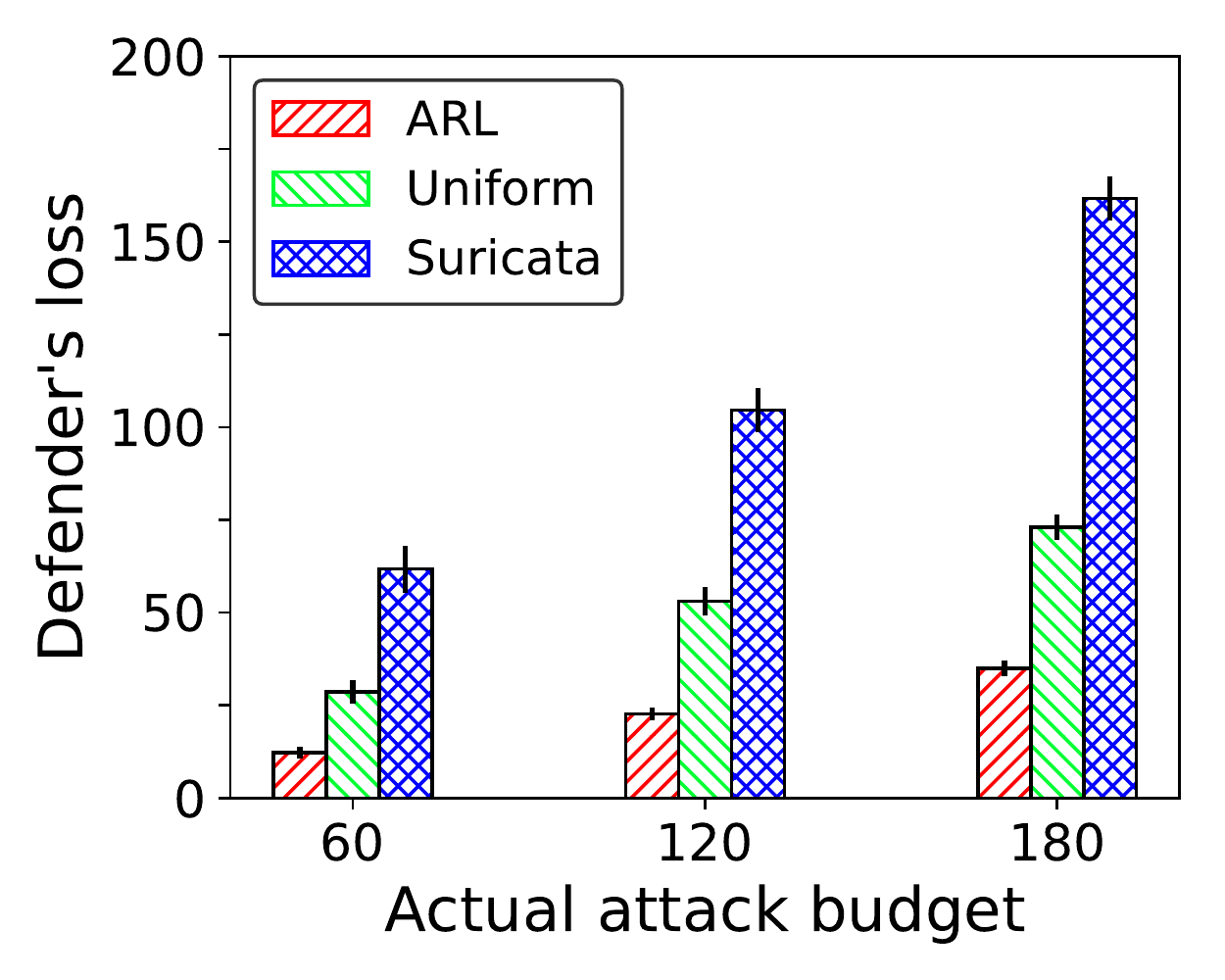} \\
\end{tabular}
\caption{Intrusion detection: loss of the defender when it is uncertain of the attack budget.  Left: def.\ budget=500.
 % Middle: def.\ budget=1000.
  Right: def.\ budget=1500. The defender's estimate of the attack budget is 120 in all cases. Thus, if the actual attack budget is 60, then the defender overestimates the adversary's budget; if the actual attack budget is 180, then it is underestimated by the defender.}
	\label{fig:ids_imcomplete}
\end{figure}

\begin{figure}[h!]
\centering
\includegraphics[width=0.22\textwidth]{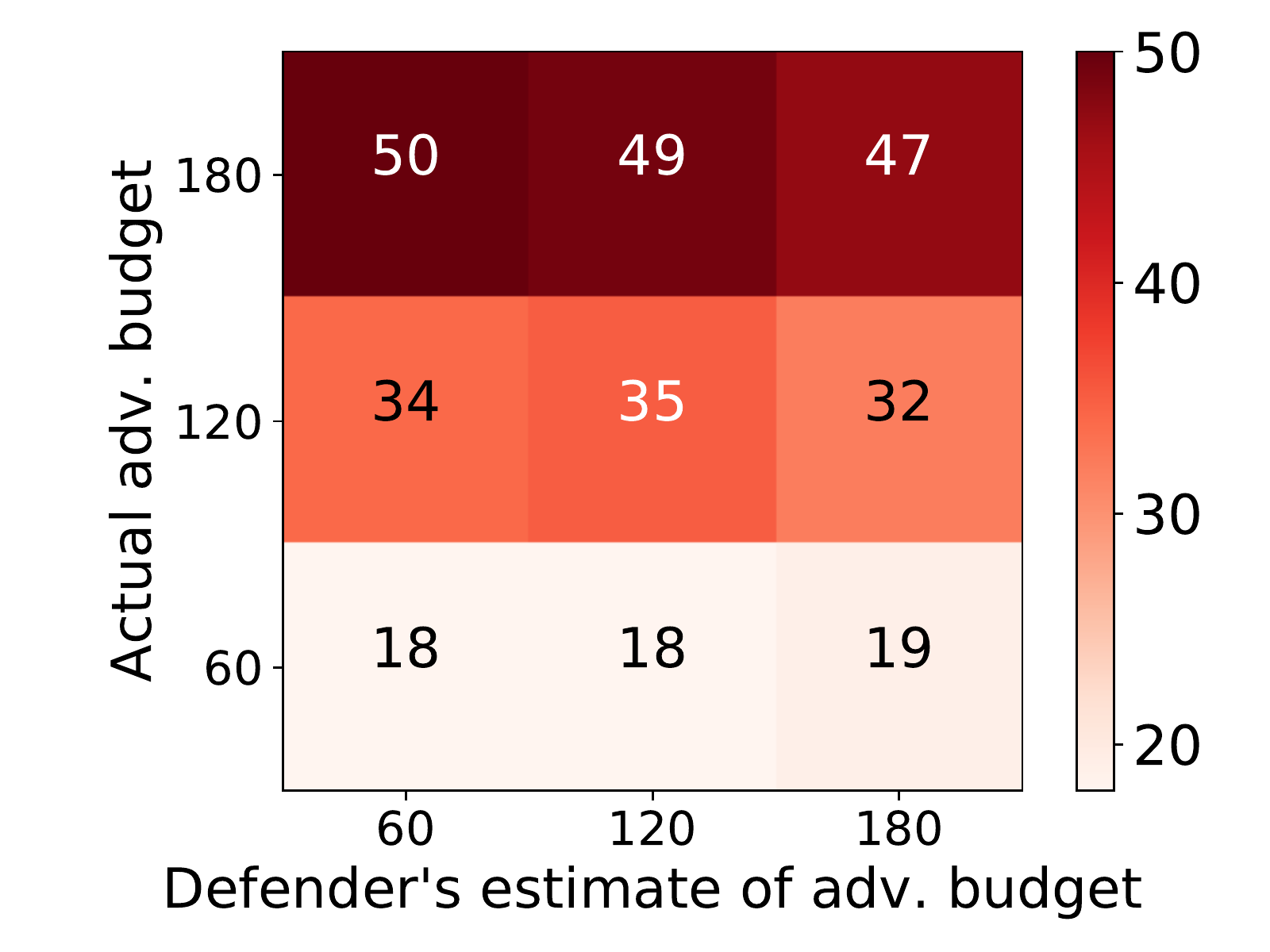}\\
\caption{Intrusion detection: loss of the defender when it has different estimates of the attack budget.}
	\label{fig:ids_cm_cross}
\end{figure}

Evaluating the alert prioritization methods when the defender is uncertain about the attack budget (Figures~\ref{fig:ids_imcomplete} and~\ref{fig:ids_cm_cross}), we can observe that the proposed ARL approach still achieves the lowest defender loss both when the attack budget is underestimated and when overestimated, and it is still far better than the baselines.
%with a loss reduction up to $50\%$ compared to Uniform, and $80\%$ compared to Suricata.
In addition, Figure~\ref{fig:ids_cm_cross} shows that when the attack budget is underestimated or overestimated, there is only a $5\%$ performance degradation compared to when the defender has full knowledge of the adversary.
This demonstrates that our approach remains robust to a strategic adversary even when the defender does not precisely know the adversary's capabilities.
Moreover, in this domain we can see that neither over- nor underestimating adversary's budget is particularly harmful, although overestimation appears to be slightly better.
%However, these results suggest that it is somewhat better to overestimate the adversary, although in this case .
%with the uncertainty of the adversary's capabilities.

\begin{figure}[h!]
\centering
\begin{tabular}{cc}
  \includegraphics[width=0.22\textwidth]{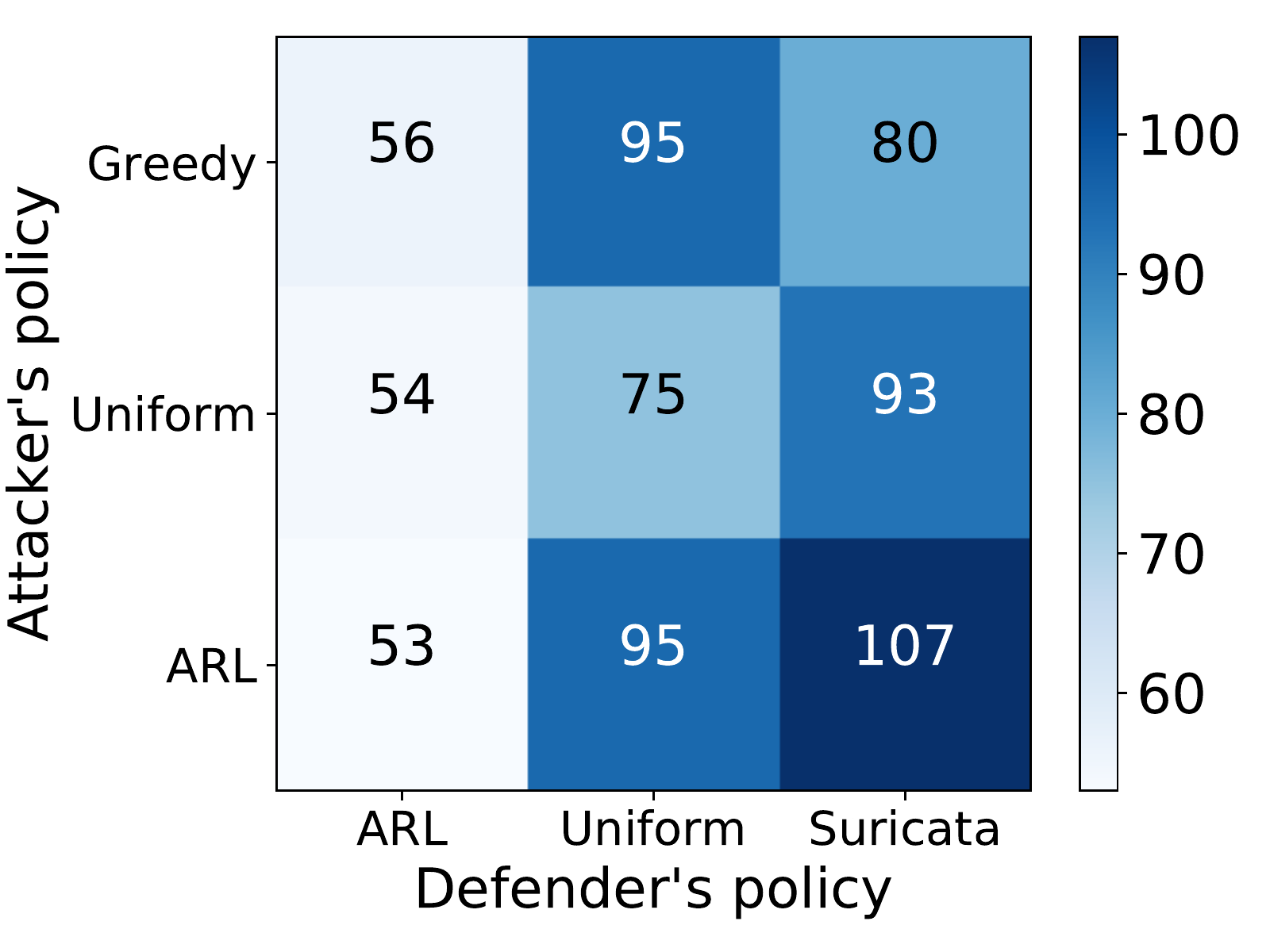} &
\includegraphics[width=0.22\textwidth]{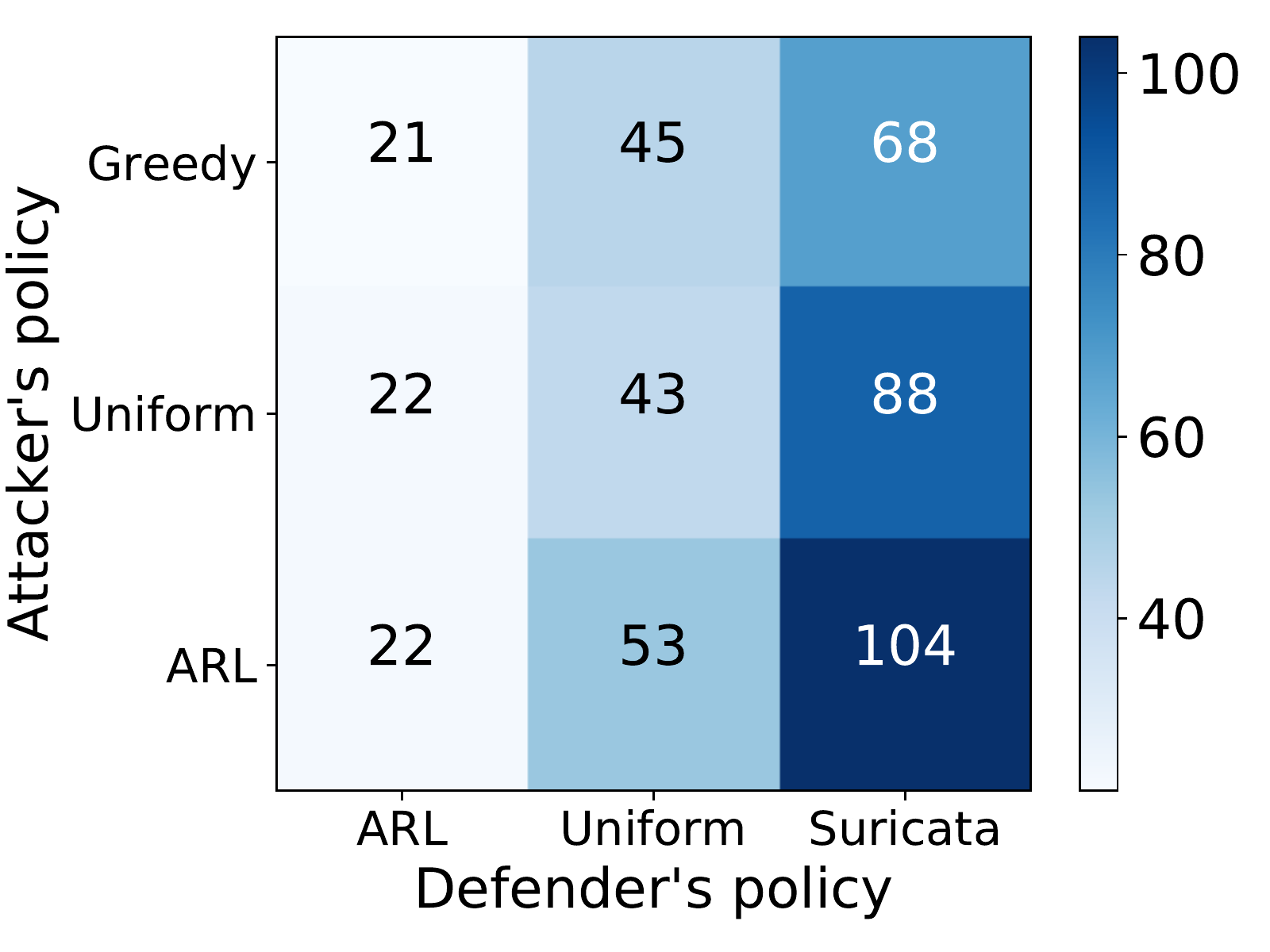} \\
\end{tabular}
\caption{Intrusion detection: loss of the defender when it is certain of the attack budget but is uncertain of the attack policy. The attack budget is fixed as 120.  Left: def.\ budget=500.
 %def\_budget=1000. (d)
  Right: def.\ budget=1500.
}
	\label{fig:ids_cm}
      \end{figure}
      Our final consideration is the impact of uncertainty about the adversary's rationality (Figure~\ref{fig:ids_cm}).
      Specifically, we now study how our approach performs, compared to the baselines, if the adversary is in some way myopic, either using a simple uniform strategy (\emph{Uniform}) or greedily choosing attacks in order of impact (\emph{Greedy}).
%Considering now the performance of alert prioritization when the adversary applies other attack polices than using RL to compute the best response to the defender (Figure~\ref{fig:ids_cm}), we also observe consistent results with the case study on fraud detection.
%First, the proposed ARL approach is robust to adversaries even with other policies (Uniform and Greedy), and it outperforms the baselines by reducing up to $80\%$ of the defender's loss.
%Second, an adversary equipped with RL can lead to the lowest defender's loss when the defense policy is given, which in turn demonstrates the adversary's preference when it has a collection of attack policies.
We can observe that although we assume a very strong adversary, our ARL approach significantly outperforms the other baselines even when the adversary is using a different attack policy.

%% file: content/case_study_fraud.tex
% We now undertake our first case study: alert prioritization in fraud detection.
While IDS settings are a natural fit for our approach, we now demonstrate its generalizability by considering a very different problem in which our goal is to identify fraudulent credit card transactions.
%Our case study relies on the learning-based fraud detector we implemented.
%From the perspective of defense, we show that the proposed approach aided by adversarial reinforcement learning provides significantly more robust alert prioritization policies compared to other baselines when the defender has, or has no knowledge of the adversary's capabilities.
Just as with the first case study, we will present the results first when the defender has full knowledge of the adversary's capabilities, and subsequently study the impact of defender's uncertainty about these.

\subsubsection{Fraud dataset}

The fraud dataset\footnote{Available at: https://www.kaggle.com/mlg-ulb/creditcardfraud.} contains 284,807 credit card transactions, of which 482 are fraudulent.
%These transactions occurred over two days by cardholders in Europe.
Each transaction is represented by a vector of 30 numerical features, 28 of which are transformed using \emph{Principle Component Analysis (PCA)}.
%thus provide no background about the values.
%The two features that are not transformed by PCA are Time and Amount: the former indicates the number of seconds between each transaction and the first transaction in the dataset, and the latter is the amount of the corresponding transaction.
In addition, each feature vector is associated with a binary label indicating the type of transaction (regular or fraudulent).
%: genuine transactions are labeled as 0, and fradulent ones have the label 1.
%if the transaction is genuine, then its label is 0; Otherwise, it has a label of 1.
In order to make it meaningful in our context, we cluster the set of fraudulent transactions into $n$ subsets, indicating a type of attack, using a Gaussian Mixture model~\cite{Bishop11}.
In our experiments, we set $n=6$, and modify the dataset with fraudulent labels replaced by cluster assignments.
%The fraud dataset is initially for binary classification, and are modified so as to represemt multiple type of frauds in our experiments.
%To do this, we use the Gaussian Mixture model to divide all the fraudulent trasactions into $n$ clusters each of which represents one type of attack action.
%We then replace the original label of each fraudulent transaction with the index of the cluster to which it is assigned.
%In our experiments, we make $n=6$.
%By doing so, the dataset is modified to have 7 classes in total, one of which indicates genuine trasactions, and the others represents 6 types of fraudulent trasactions.
The counts of each type of transaction is shown in Table \ref{tab:fraud_type}. 

\begin{table}[h!]
\caption{Number of transactions in the modified fraud~dataset}
\centering
\begin{tabular}{|c|c|c|}
\hline
\textbf{Original transaction type}   & \textbf{Label} & \textbf{Count}   \\ \hline\hline
Genuine                     & 0     & 284,308 \\ \hline
\multirow{6}{*}{Fraudulent} & 1     & 11      \\ \cline{2-3} 
                            & 2     & 21      \\ \cline{2-3} 
                            & 3     & 72      \\ \cline{2-3} 
                            & 4     & 250     \\ \cline{2-3} 
                            & 5     & 14      \\ \cline{2-3} 
                            & 6     & 124     \\ \hline
\end{tabular}
\label{tab:fraud_type}
\end{table}

\subsubsection{Learning-based fraud detector}

We developed a fraud detector using supervised learning on the fraud dataset.
%A conventional approach is to use multi-class classification.
The main challenge is that the dataset is highly imbalanced, as shown in Table \ref{tab:fraud_type}: the fraudulent transactions only account for $<0.2\%$ of all transactions.
%, let alone each type of fraudulent transaction.
To address this challenge, we apply \emph{Synthetic Minority Over-sampling Technique (SMOTE)} to produce synthetic data for the minority classes to balance the data.
Our implementation contains the following steps:

%\begin{enumerate}[label=(\roman*)]

%\item \emph{Dataset splitting:}
(i) \textit{Dataset splitting}: We use stratified split to partition the modified fraud dataset into training and test data with equal size, which contain roughly the same proportions of the fraudulent and non-fraudulent data.

%\item \emph{Binary classification:}
(ii) \textit{Binary classification:} We use SMOTE and linear SVM to learn a binary classifier to predict whether a transaction is fraudulent.
  %\footnote{Here we consider that the dataset contains labels of zero and nonzero thus our task is binary classification at this stage.}
%This can be done by using SMOTE to oversample the fraudulent data in the training set, then training a linear SVM classifier on the balenced training dataset.  
The resulting classifier has an AUC $>$99\% and a recall $>$90\% on the test data, which indicates that more than $90\%$ of the fraudulent transactions can be detected. 

%\item \emph{Multi-class classification:}
(iii) \textit{Multi-class classification:} We now restrict attention to only the fraudulent transactions to learn a conditional classifier to predict the type of fraud.
%  only on the fraudulent transactions in the training set and build a collection of classifiers to predict the fraud types.
Specifically, we learn 6 independent classifiers each of which corresponds to one fraud type and returns a binary classification result indicating whether a fraudulent transaction belongs to this type.  
%If the answer is affirmative, then an alert corresponding to this type of fraud is triggered.
Similarly to Step (ii), we use SMOTE and linear SVM to learn these classifiers, each of which admits $>94\%$ recall.
%This indicates that when only fraudulent transactions are considered, each one is predicted with its correct type with a probability $>$94\%. 
%
%\end{enumerate}

Once the fraud detector is implemented, we evaluate the detector using the test dataset.
We first predict the test data by using the binary classifier obtained in Step (ii) above.
If any transaction in the test data is classified as fraudulent, then it is further inspected by the 6 classifiers we construct for multi-class classification. 
If a fraudulent transaction is predicted as any type of fraud, then a corresponding alert is triggered.
Otherwise, an alert corresponding to the fraud type which is predicted with the highest classification score is triggered.
%In other words, if a transaction is predicted as fraudulent by the binary classifier at the first stage, it triggers at least one type of alert.

\subsubsection{Experimental Setup}

To evaluate the robustness of the proposed approach for alert prioritization in fraud detection, we first computed the distributions of the true and false alerts identified by the fraud detector that we implemented.
By doing so, we obtained the probability that any attack $a\in A$ triggers an alert $t\in T$, as well as the number of false alarms associated with each type of alert, each of which has a value of 1 as the investigation cost.
We filtered out alert types that were triggered \emph{only} by fraudulent transactions (as we had done before), leaving 3 out of 6 alert types. 
We also filtered out the attack actions which are associated with the alert types omitted above, as these attacks can always be detected by investigating the corresponding alerts.
The resulting distribution of the alerts triggered by frauds is given in Table~\ref{tab:fraud_attack_alert}.
\begin{table}[h!]
\caption{Probability that an attack action triggers each type of alert}
\centering
\scalebox{0.85}{
\begin{tabular}{|c|c|c|c|}
\hline
\multirow{2}{*}{\textbf{Attack action}} & \multicolumn{3}{c|}{\textbf{Alert type}} \\ \cline{2-4} 
                                      & 1            & 2           & 3           \\ \hline\hline
1                                     & 0.9          & 0.61        & 0           \\ \hline
2                                     & 0.09         & 0.87        & 0.12        \\ \hline
3                                     & 0            & 0.41        & 0.85        \\ \hline
\end{tabular}
}
\label{tab:fraud_attack_alert}
\end{table}

We used $[1,3,2]$ as the adversary's cost of the mounting each type of attack action.
We employed the mean amount of each type of fraudulent transaction as the loss of the defender if any such type of attack action is not detected, measured by the unit of 10 Euros.
The corresponding defender's loss for each undetected attack was $[9.4, 12.1, 16.0]$.
In addition, we used 30 minutes as the fixed length of each time period in our experiments. 
Based on our classification results, the average number of false alerts that occur of each type in a time period was $[10,47,39]$. 
Similar to our IDS case study, we simulated the distribution of false alerts by using Poisson processes with the above mean values.
%\footnote{Due to the limited samples of the number of false alerts raised in each time period, we were unable to conduct a specific distribution fitting for each type of false alerts.
%Here we use the Poisson distribution to approximate the actual pattern.} 

\subsubsection{Baselines}

The performance of the proposed approach is investigated by comparing with three alternative policies for alert prioritization:
\emph{Uniform}, a policy which uniformly allocates the defender's budget over each alert type;
\emph{GAIN}~\cite{laszka2017}, a game theoretic approach which prioritizes alert types, and always inspects all alerts of a selected type; and
% and exhausts the defender's budget according to the prioritization;
\emph{RIO}~\cite{yan2018}, another game theoretic approach which prioritizes alerts, and computes an approximately optimal number of alerts of each type to inspect.
%and determines a threshold for the maximum budget that the defender can distribute on each type of alerts.

\subsubsection{Results}
\label{sec:case_study_fraud}

\begin{figure}[h!]
\centering
\begin{tabular}{cc}
  \includegraphics[width=0.22\textwidth]{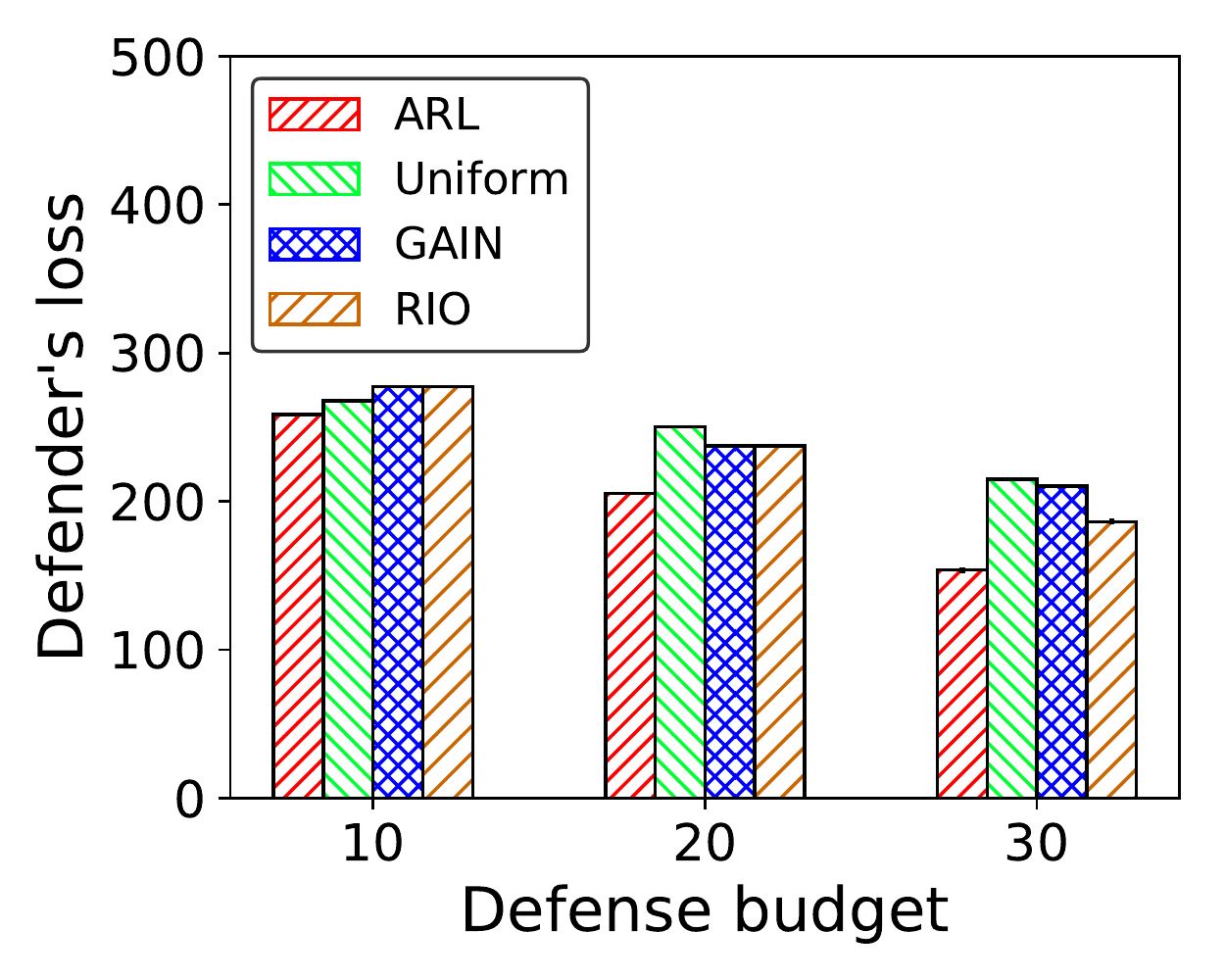} &
 \includegraphics[width=0.22\textwidth]{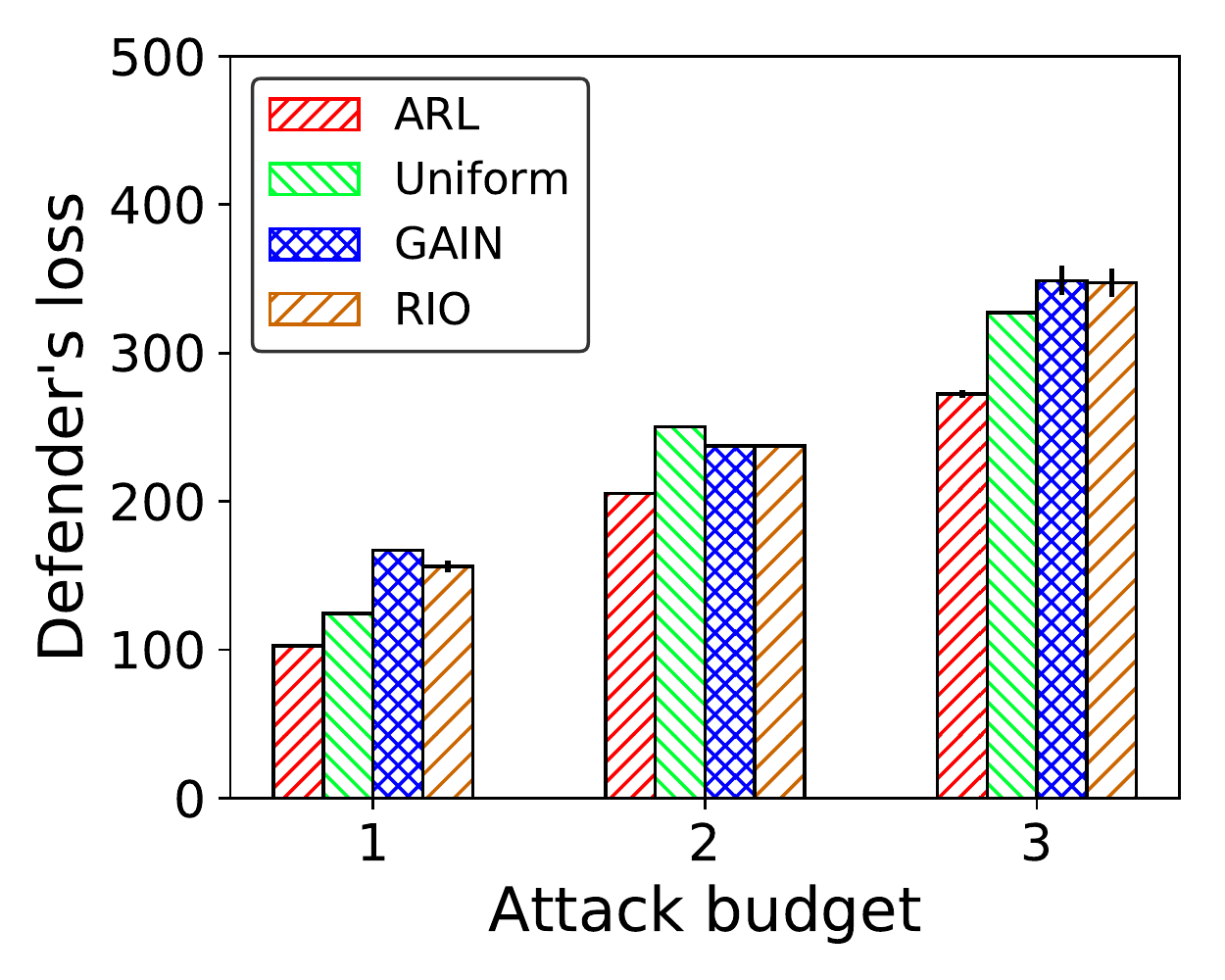}\\
\end{tabular}
	\caption{Fraud detection: loss of the defender when it knows the attack budget.  Left: Defender's loss by its budget, with attack budget adv\_budget being fixed as 2.  Right: Defender's loss by attack budget, with defense budget def\_budget being fixed as 20.}
	\label{fig:fraud_complete}
\end{figure}

Figure~\ref{fig:fraud_complete} shows the results when the defender has full knowledge of the adversary's capabilities.
%There are several findings that we wish to highlight.
We can observe that the proposed approach (\emph{ARL}) outperforms other baselines in all settings, typically by at least $25\%$.
The main reason for the advantage is similar to that in the IDS setting: the ability to have a policy that is carefully optimized and conditional on state significantly increases its efficiency.
%with a $25\%$ reduction on the defender's loss.
%This demonstrates that by assuming a strong adversary model and strategically prioritizing alerts which is adaptive to the observed states, the loss of the defender can be dramatically reduced. 
%Second, the defender's loss is affected by the defense budget (termed as \emph{def\_budget}) and attack budget (termed as \emph{adv\_budget}).
%For all our settings, larger amounts of def\_budget and smaller amounts of adv\_budget lead to less defender's loss.
%When the defense budget increases to 30, or the attack budget decreases to 1, the loss reduction compared to other baselines can be up to $50\%$.
Interestingly, the alternative game theoretic alert prioritization approaches, GAIN and RIO, are in some cases worse than the uniformly random policy.
The key reason is that they can be myopic in that they independently optimize for a single time period, whereas attacks can be adaptive.
The proposed approach, in contrast, explicitly considers such adaptivity.
%the two baseline methods based on security games, can even fall behind the naive Uniform policy in terms of robustness.
%As shown in Figure~\ref{fig:fraud_complete}, when def\_budget decreases to 10, or adv\_budget increases to 3, the Uniform policy leads to slightly less defender's loss than its counterparts using security games.
%The major reason is that GAIN and RIO  assume that alert prioritization is chosen before observing the actual number of alerts.
%In addition, they design the games by considering a single time period, while the actual adversary can mount attacks adaptively during consecutive time periods.

\begin{figure}[h!]
\centering
\begin{tabular}{cc}
  \includegraphics[width=0.22\textwidth]{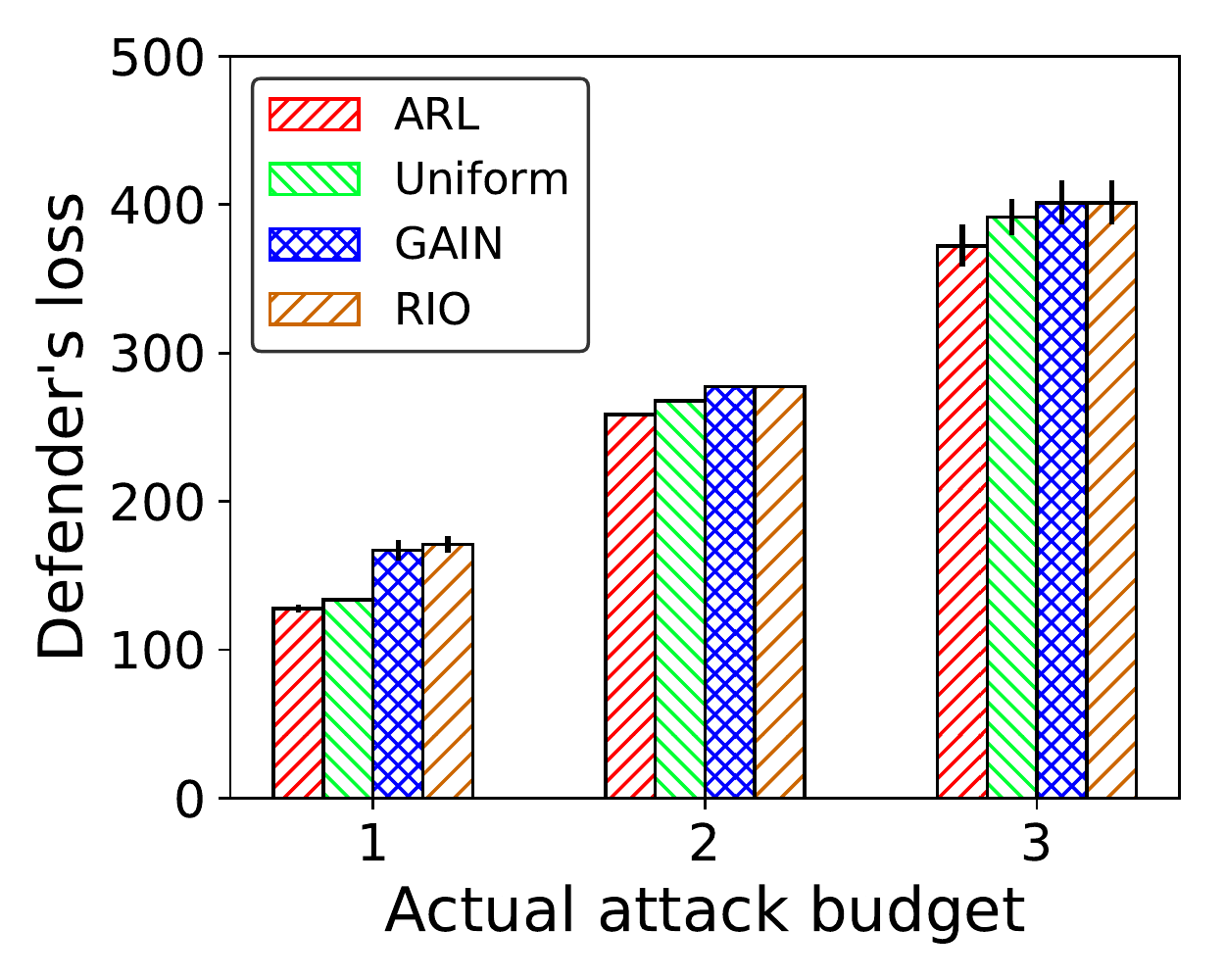} &
 \includegraphics[width=0.22\textwidth]{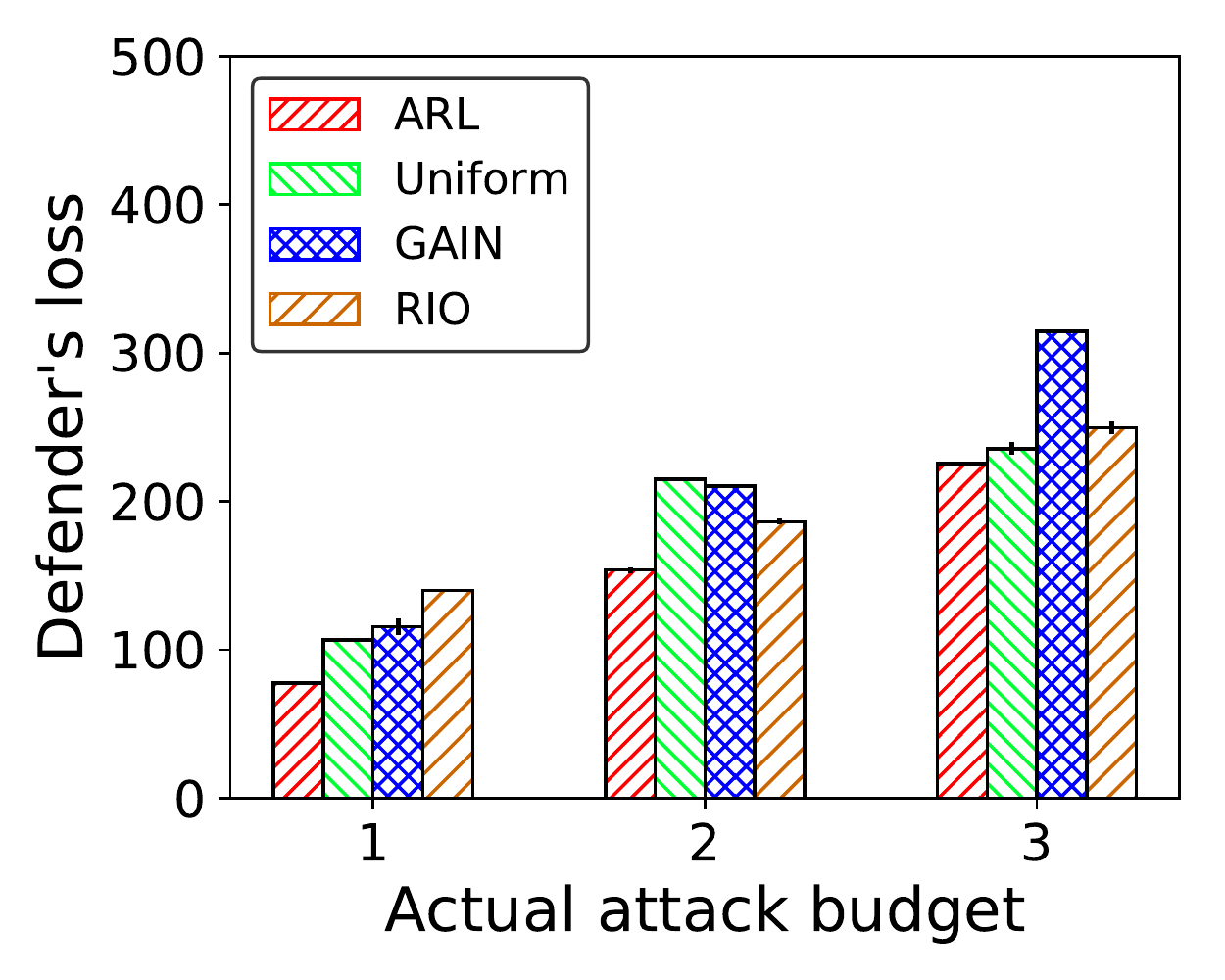} \\
\end{tabular}
\caption{Fraud detection: loss of the defender when it is uncertain of the attack budget.  Left: def.\ budget=10.
 % Middle: def\_budget=20.
  Right: def.\ budget=30. The defender's estimate of the attack budget is 2. If the actual attack budget is 1, then the defender overestimates the adversary's budget; if the actual attack budget is 3, then it is underestimated.}
	\label{fig:fraud_imcomplete}
      \end{figure}
      \begin{figure}[h!]
\centering
\includegraphics[width=0.22\textwidth]{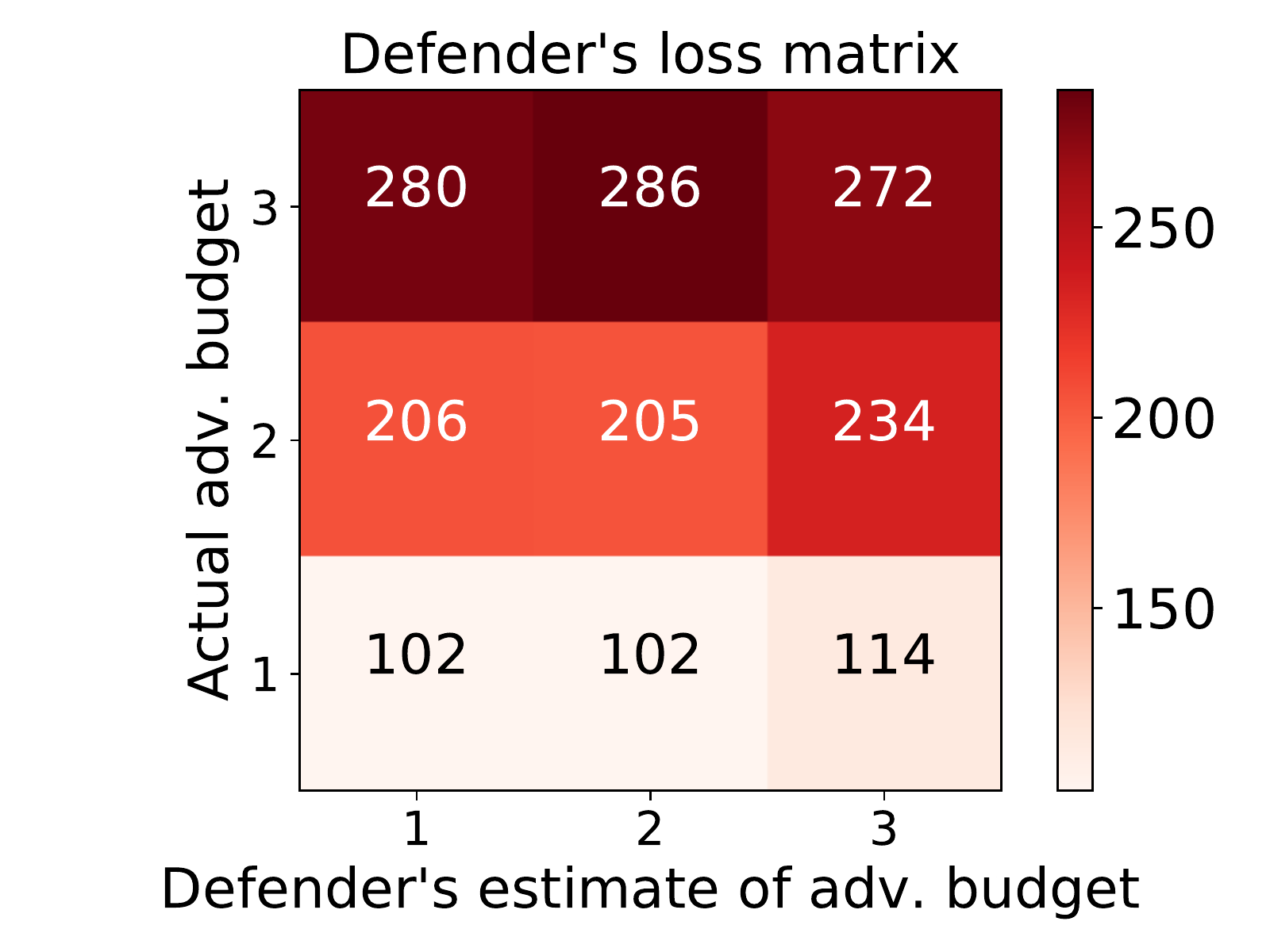}\\
\caption{Fraud detection: loss of the defender when it has different estimates of the attack budget.}
	\label{fig:fraud_cm_cross}
\end{figure}
Figures \ref{fig:fraud_imcomplete} and \ref{fig:fraud_cm_cross} investigate performance of our approach when the attack budget is uncertain.
%present the  robustness of diffent alert prioritization polices when the defender has no knowledge of the attack budget.
It can be seen in Figure~\ref{fig:fraud_imcomplete} that ARL remains the best approach to use, despite this uncertainty.
Interestingly, \emph{GAIN} can, in contrast, be rather fragile to such uncertainty.
%still lead to the least loss of the defender compared to the other baselines, even in the cases when the attack budget is underestimated or overestimated.
%Besides, Figure \ref{fig:fraud_imcomplete} (right) shows that the reduction of defender's loss compared to other methods is the maximal (up to 50\%) when def\_budget = 30, the estimated adv\_budget = 2 and the actual adv\_budget = 1.
%This implies that the performance gap between ARL and others is maximized when there is a strong defender and weak adversary, and the adversary's capability is overestimated.
Considering Figure \ref{fig:fraud_cm_cross}, both under- and overestimation of the attack budget incurs a limited performance impact ($<10\%$).
More interesting, however, is the observation that it is actually better to slightly \emph{underestimate} the adversary's budget: in the worst case, this hurts performance less than $3\%$.
Effectively, the approach remains quite robust even against stronger attacks, whereas overestimating the budget does not take sufficient advantage of weaker adversaries.
%shows that both underestimation and overestimation lead to a reduction of the performance, but the reduction is quite limited (less than $10\%$).

\begin{figure}[h!]
\centering
\begin{tabular}{cc}
  \includegraphics[width=0.22\textwidth]{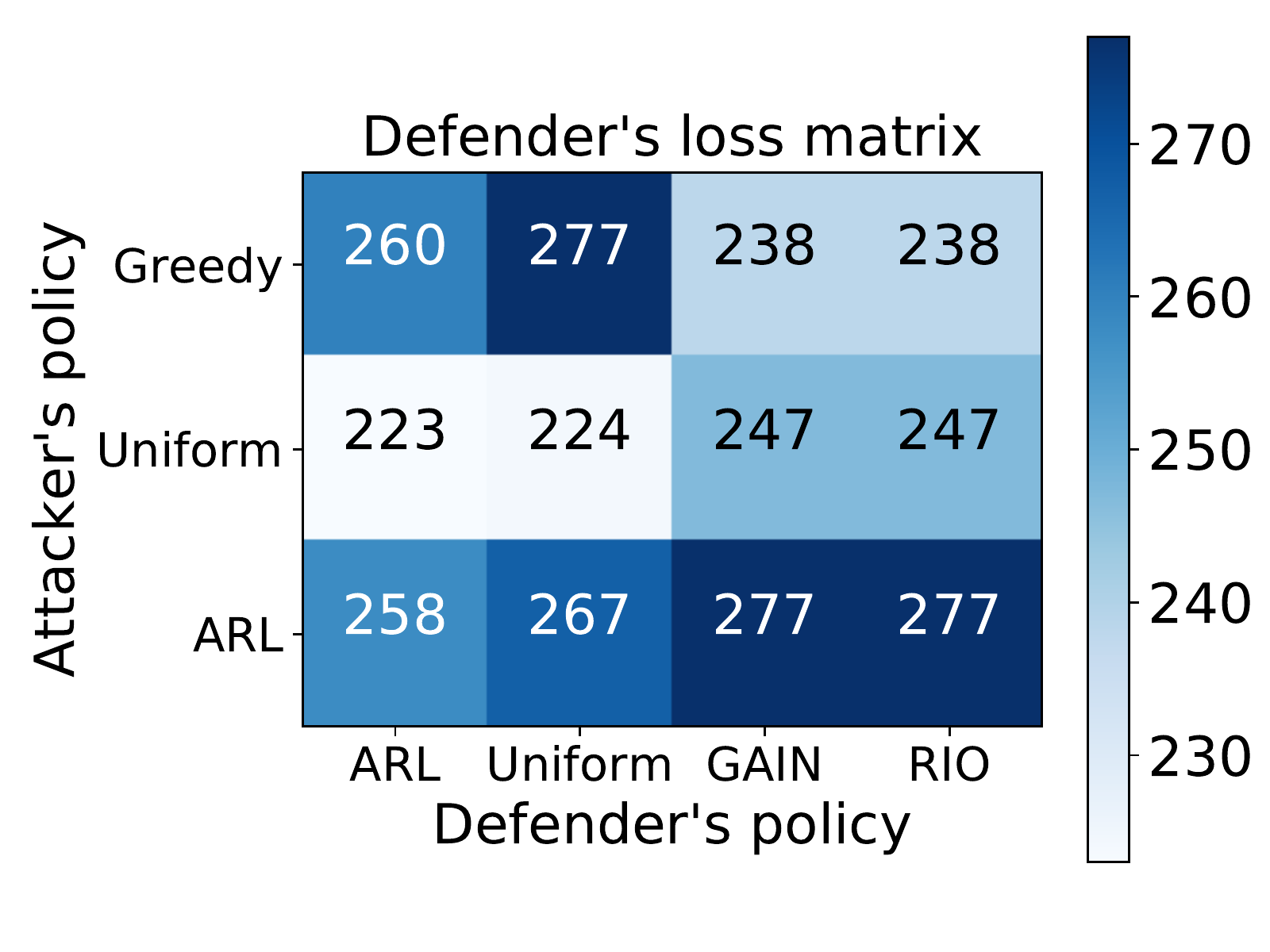} &
 \includegraphics[width=0.22\textwidth]{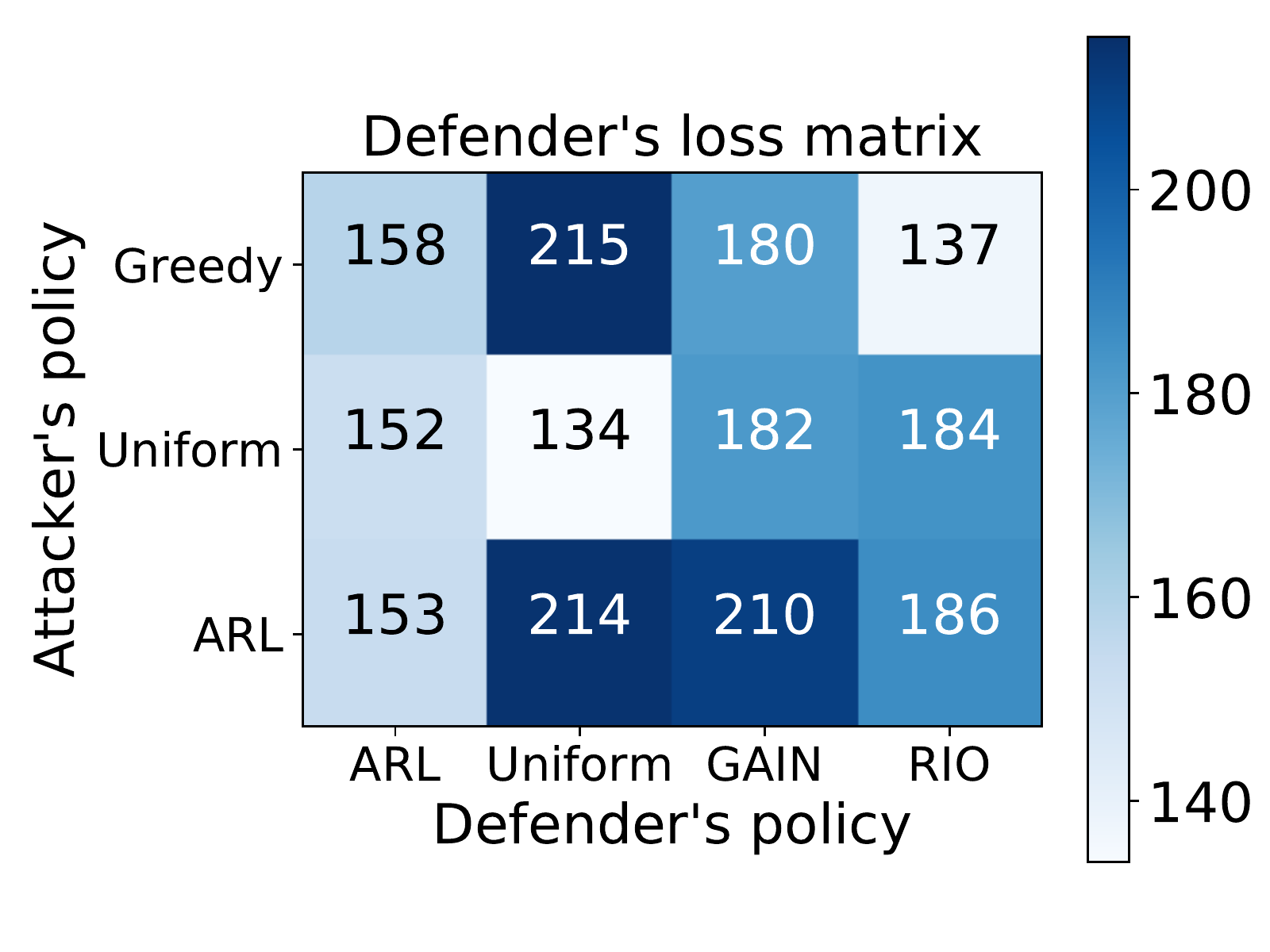} \\
\end{tabular}
\caption{Fraud detection: loss of the defender when it is certain of the attack budget but is uncertain of the attack policy. The attack budget is fixed as 2.  Left: def.\ budget=10.
%  Middle: def\_budget=20.
  Right: def.\ budget=30.}
	\label{fig:fraud_cm}
      \end{figure}
      Finally, we study the robustness of ARL compared to other baselines when the attacker is using different policies (\emph{Uniform} or \emph{Greedy}) instead of the RL-based policy that is assumed by our approach (Figure~\ref{fig:fraud_cm}).
      Here, the results are slightly more ambiguous than we observed in the IDS domain: when the adversary is using the \emph{Greedy} policy, \emph{RIO} does outperform ARL by $8\%$ when the defender's budget is small, and by $13\%$ when the defender's budget is large.
      However, in these cases, the adversary can gain a great deal by more carefully designing its policy.
      Thus, when the defender's budget is large, a rational adversary can cause \emph{RIO} to degrade by nearly $18\%$, where ARL is quite robust to such adversaries.
%Figure \ref{fig:fraud_cm} displays the robustness of alert prioritization approaches when the defender knows the attack budget but is unknown about actual policy taken by the adversary. 
%It can be seen that the proposed ARL approach leads to the optimal, or near-optimal defense (with less than $10\%$ performance degradation than the optimal one) against different attack policies.
%On the other hand, Figure \ref{fig:fraud_cm} shows that an adversary which uses RL to compute its best response to the defender outperforms the alternative policies in most of the cases, leading to up to $50\%$ more loss of the defender, which in turn validates our assumption that the adversary uses RL as its attack policy. 

%% file: content/related_work.tex
\section{Related Work}
\label{related_work}

\subsection{Deep Reinforcement Learning}

Reinforcement learning has received significant attention in recent years, which is in large part due to the emergence of deep reinforcement learning.
Deep reinforcement learning combines classic reinforcement learning approaches, such as \emph{Q-learning}~\cite{watkins1989learning}, with deep neural networks.
Classic Q-learning is a model-free reinforcement learning approach, which is guaranteed to find an optimal policy for any finite Markov decision process~\cite{watkins1992q}.
However, to do so, it needs to learn and store an exact representation of the action-value function, which is infeasible for a problem with large action or state spaces.
Notable early successes combining reinforcement learning with neural networks include \emph{TD-Gammon}, a backgammon program that achieved a level of play that was comparable to top human players in 1992~\cite{tesauro1994td}.
% DeepMind ATARI
% ``Playing Atari with deep reinforcement learning''
% \url{https://arxiv.org/pdf/1312.5602.pdf}
% ``Human-level control through deep reinforcement learning''
% \url{https://www.nature.com/articles/nature14236}
More recently, Mnih et al.\ introduced the model-free \emph{Deep Q-Learning algorithm} (DQN), which achieved human-level performance in playing a number of Atari videogames, using purely visual input from the games~\cite{mnih2013playing,mnih2015human}.
However, the actions spaces in all of these games were small and discrete.
% DDPG
% ``Continuous control with deep reinforcement learning''
% \url{https://arxiv.org/abs/1509.02971}
Lillicrap et al.\ adapted the idea of Deep Q-Learning to continuous action spaces by introducing an algorithm, called \emph{Deep Deterministic Policy Gradient} (DDPG)~\cite{lillicrap2016}.
DDPG is a model-free actor-critic algorithm, whose robustness is demonstrated on a variety of continuous control tasks.
% Rainbow
% ``Rainbow: Combining Improvements in Deep Reinforcement Learning''
% \url{https://arxiv.org/pdf/1710.02298.pdf}
Hessel et al.\ evaluated six improvements to the DQN algorithm (DDQN~\cite{van2016deep}, Prioritized DDQN~\cite{schaul2015prioritized}, Dueling DDQN~\cite{wang2016dueling}, A3C~\cite{mnih2016asynchronous}, Distributional DQN~\cite{bellemare2017distributional}, and Noisy DQN~\cite{fortunato2017noisy}), which had been proposed by the deep reinforcement learning community since the publication of DQN, across 57 Atari games~\cite{hessel2018rainbow}.
Further, they integrated these improvements into a single agent, called \emph{Rainbow}, and demonstrated its state-of-the-art performance on common benchmarks.

\subsection{Multi-agent Reinforcement Learning}

Single-agent reinforcement learning approaches can train only one agent at a time, which means that in a multi-agent setting, they must treat other agents as part of the environment.
As a result, they often provide policies that are not robust---especially in a non-cooperative setting such as ours---since they cannot consider the possibility that other agents respond by learning and updating their own policies.
Multi-agent reinforcement learning approaches attempt to provide more robust policies by training multiple adaptive agents together.

% minimax Q
% ``Markov games as a framework for multi-agent reinforcement learning''
% \url{https://www.sciencedirect.com/science/article/pii/B9781558603356500271}
Littman proposed a framework for multi-agent reinforcement learning that models the competition between two agents as a zero-sum Markov game~\cite{littman1994markov}.
To solve this game, the author introduced a Q-learning-like algorithm, called \emph{minimax-Q}, which is guaranteed to converge to optimal policies for both players.
However, the minimax-Q algorithm assumes that the game is zero-sum (i.e., the player's rewards are exact opposites of each other) and every step of the training involves exhaustive searches over the action spaces, which limits the applicability of the algorithm.
A number of follow up efforts have proposed more general solutions.
% ``Nash Q-Learning for General-Sum Stochastic Games''
% \url{http://www.jmlr.org/papers/volume4/hu03a/hu03a.pdf}
For example,
Hu and Wellman extended Littman's framework to general-sum stochastic games~\cite{hu1998multiagent}.
They propose an algorithm that is based on each agent learning two action-value functions (one for itself and one for its opponent), which is guaranteed to converge to a Nash equilibrium under certain conditions.
To relax some of these conditions, Littman introduced \emph{Friend-or-Foe Q-learning}, in which each agent is told to treat each other agent either as a ``friend'' or as a ``foe''~\cite{littman2001friend}.
Later, Hu and Wellman proposed the \emph{NashQ algorithm}, which generalizes single-agent Q-learning to stochastic games with many agents by using an equilibrium operator instead of expected utility maximization~\cite{hu2003nash}.

While the above approaches have the advantage of providing certain convergence guarantees, they assume that action-value functions are represented exactly, which is infeasible for scenarios with large action or state spaces.
Deep multi-agent reinforcement-learning provides a more scalable approach by representing action-value functions using deep neural networks.
% ``Multi-Agent Actor-Critic for Mixed Cooperative-Competitive Environments''
% \url{https://arxiv.org/abs/1706.02275}
For example, 
Lowe et al.\ proposed an adaptation of actor-critic reinforcement-learning methods to multi-agent settings~\cite{lowe2017multi}.
In the proposed approach, each agent learns a collection of different sub-policies, and for each episode, each agent randomly selects sub-policy from this collection.
However, in contrast to our approach, the size of the collection is fixed (which may waste training effort at the beginning and might not converge in the end) and the agents choose their sub-policies at random instead of strategically.
%
% policy-space response oracles (i.e., generalization of double oracle)
% ``A Unified Game-Theoretic Approach to Multiagent Reinforcement Learning''
% \url{https://arxiv.org/pdf/1711.00832.pdf}
Lanctot et al.\ introduced an algorithm, called \emph{policy-space response oracles}, which is closer to our double-oracle based computational approach~\cite{lanctot2017}.
Their proposed algorithm maintains a set of policies for each agent, but it does not incorporate actor-critic methods, and it was evaluated in settings with relatively small discrete action spaces.

\subsection{Alert Management and Prioritization}

A multitude of research efforts have studied the problem of reducing the number of alerts without significantly reducing the probability of attack detection~\cite{hubballi2014false}.
One of the most common approaches is \emph{alert correlation} and \emph{clustering}, which attempt to group related alerts together, thereby reducing the set of messages that are presented~\cite{salah2013model}.
In distributed systems, \emph{collaborative intrusion detection systems} may be deployed, which include several monitoring components and correlate alerts among the monitors to create a holistic view~\cite{vasilomanolakis2015taxonomy}.
Since the number of alerts may be too high even after correlation, research efforts have also investigated the prioritization of alerts.
For example, Alsubhi et al.\ introduced a fuzzy-logic based alert management system, called \emph{FuzMet}, which uses several metrics and fuzzy logic to score and prioritize alerts~\cite{alsubhi2012fuzmet}.
However, these approaches do not consider the possibility of an attacker adapting to the prioritization.

\subsection{Game Theory for Alert Prioritization and Security Audits}

Prior work has successfully applied game theory to a variety of security problems, ranging from physical security~\cite{an2013deployed} to network security and privacy~\cite{manshaei2013game}.

% TODO: brown2016 ?

Our approach is most closely related to \emph{alert-prioritization games}.
% our AICS'17 paper
Laszka et al.\ introduced the first game-theoretic model for alert prioritization, which they solved with the help of a greedy heuristic~\cite{laszka2017}.
The performance of this approach, which we denoted GAIN in our experiments, %(Section~\ref{sec:baseline_approaches}),
is limited by its restrictive assumptions about the defender's decision making.
In particular, GAIN assumes that the defender's policy is a strict prioritization that investigates all higher-priority alerts before investigating any lower-priority ones, and the prioritization is chosen before observing the actual number of alerts.
Moreover, the model considers only a single time slot, which further limits its usefulness.
% our ICDE'18 paper
Yan et al.\ improved upon GAIN by allowing the defender to specify a maximum budget that may be spent on each alert types, thereby relaxing the strict prioritization of GAIN~\cite{yan2018}.
However, this improved approach, which we denoted RIO in our experiments, %(Section~\ref{sec:baseline_approaches}),
still assumes that the prioritization is chosen before observing any alerts and considers only a single time slot.
As our numerical results demonstrate, these restrictions can lead to significantly higher losses for the defender.
% TEAMCORE paper (IJCAI'17)
Schlenker et al.\ introduced a similar model, called \emph{Cyber-alert Allocation Game}, which further simplifies the problem by assuming that the number of false alerts is fixed and known by both parties in advance~\cite{schlenker2017}.

Our approach also resembles \emph{audit games}, which study the problem of allocating a limited amount of audit resources to a fixed number of audit targets~\cite{blocki2013,blocki2015audit}.
However, despite the resemblance, audit games are ill-suited for prioritizing alerts since these games assume that the attacker knows the exact set of targets, which would correspond to individual alerts, before launching its attack.
Due to the unpredictability of false alerts, this assumption does not hold for alert prioritization.

%% file: content/conclusion.tex
\section{Discussion and Conclusion}
\label{conclusion}

% REITERATRE MOTIVATION AND IMPORTANCE OF PROBLEM

% LONGER VERSION
%Since there is generally a tradeoff between reducing the rate of false alarms and improving detection probability, sensitive detection systems tend to generate myriads of false alerts. 
%Even after applying techniques for reducing the alert burden (e.g., alert correlation), there often remain vastly more alerts than time to investigate them, which means that the success of detection often hinges on how defenders prioritize certain alerts over others.
% SHORTHER VERSION
Since even after applying techniques for reducing the alert burden (e.g., alert correlation) there often remain vastly more alerts than time to investigate them, the success of detection often hinges on how defenders prioritize certain alerts over others.
In practice, prioritization is typically based on non-strategic heuristics (e.g., Suricata's built-in priority values), which may easily be exploited by a strategic attacker who can adapt to the prioritization.
Strategic prioritization approaches attempt to prevent this by using game-theory to capture adaptive attackers; however, existing strategic approaches severely restrict the defender's policy (e.g., strict prioritization) for the sake of computational tractability.

% REITERATRE KEY ACHIEVEMENTS

In contrast, we introduced a general model of alert prioritization that does not impose any restrictions on the defender's policy, and we proposed a novel double oracle and reinforcement learning based approach for finding approximately optimal prioritization policies efficiently.
Our experimental results---based on case studies of IDS and fraud detection---demonstrate that these policies significantly outperform non-strategic prioritization and prior game-theoretic approaches.
Further, to demonstrate the strength of our attacker model, we also showed that the attacker policies found by our approach outperform multiple baseline policies.

% APPLICABILITY AND FUTURE WORK

For practitioners, the key task in applying our approach is estimating the parameter values of our model.
In our case studies, we showed how to estimate parameters in two domains (e.g., for IDS, using CVSS score to estimate attack impact and CVSS complexity for attack cost).
The most difficult parameter to estimate is the attacker's budget; however, our experimental results show that our approach is robust to uncertainty in the attacker's budget and outperforms other approaches even when the budget is misestimated.
We leave studying the sensitivity to other parameters to future work.

%% file: content/appendix.tex
\appendix

\renewcommand{\algorithmicrequire}{ \textbf{Input:}} %Use Input in the format of Algorithm  
\renewcommand{\algorithmicensure}{ \textbf{Output:}} %UseOutput in the format of Algorithm 

\subsection{Best Response Oracle Algorithm}
The proposed algorithm to compute the best response oracle is outlined in Algorithm~\ref{alg:ddpg_mix}.

\begin{algorithm}[h!]   
\caption{DDPG-MIX Algorithm: Compute the pure-strategy best response of player $v$ when its opponent takes mixed-strategy $\bm{\sigma}_{-v}$.}   
\label{alg:ddpg_mix}   
\begin{algorithmic}[1]  
\REQUIRE ~~\\
The set of opponent's pure strategies, $\bm{\Pi}_{-v}$; \\
Mixed strategy of the opponent, $\bm{\sigma}_{-v}$; 
\ENSURE ~~\\
The value network of player $v$, $Q_{v}(\bm{O}_v, \bm{\alpha}_v|\bm{\theta}_v^Q)$; \\   
The policy network of player $v$, $\bm{\pi}_{v}(\bm{O}_v|\bm{\theta}_v^\pi)$;
\STATE Randomly initialize $Q_{v}(\bm{O}_v, \bm{\alpha}_v|\bm{\theta}_v^Q)$ and $\bm{\pi}_{v}(\bm{O}_v|\bm{\theta}_v^\pi)$;
\STATE Initialize replay memory $\mathcal{D}$;
\FOR{$episode = 0,M-1$}
\STATE Initialize the system state $\langle \bm{N}^{(0)}, \bm{M}^{(0)}, \bm{S}^{(0)} \rangle = \langle \bm{0}, \bm{0}, \bm{0} \rangle$; 
\STATE Sample the opponent's policy $\bm{\pi}_{-v}$ by using $\bm{\sigma}_{-v}$ over $\bm{\Pi}_{-v}$;
\FOR{$k = 0, K-1$}
\STATE With probability $\epsilon$ select a random action $\bm{\alpha}_v^{(k)}$; Otherwise, select $\bm{\alpha}_v^{(k)}=\bm{\pi}_{v}(\bm{O}_v^{(k)}|\bm{\theta}_v^\mu)$;
\STATE Execute $\bm{\alpha}_v^{(k)}$ and $\bm{\alpha}_{-v}^{(k)}=\bm{\pi}_{-v}(\bm{O}_{-v}^{(k)})$, observe reward $r_v^{(k)}$ and transit the system state to $\bm{S}^{k+1}$;
\STATE Store transition $(\bm{O}_{v}^{(k)}, \bm{\alpha}_v^{(k)}, r_v^{(k)}, \bm{O}_{v}^{(k+1)})$ in $\mathcal{D}$;
\STATE Sample a random minibatch of $N$ transitions $(\bm{O}_{v}^{(i)}, \bm{\alpha}_v^{(i)}, r_v^{(i)}, \bm{O}_{v}^{(i+1)})$ from $\mathcal{D}$;
\STATE Set $y_v^{(i)}=r_{v}^{(i)}+\tau Q_v(\bm{O}_v^{(i+1)}, \bm{\pi}(\bm{O}_v^{(i+1)}|\bm{\theta}^\pi_v)|\bm{\theta}^Q_v)$;
\STATE Update the value network by minimizing the loss 
\[ \mathcal{L}(\bm{\theta}^Q_v) = \frac{1}{N}\sum_i (y_v^{(i)} - Q_v(\bm{O}_v^{i}, \bm{\alpha}_v^{(i)}|\bm{\theta}^Q_v))^2; \]
\STATE Update the policy network by using the sampled policy gradient:
%\[ \nabla_{\bm{\theta}_v^\pi} J \approx \frac{1}{N} \sum_i \nabla_{\bm{\alpha}_v} Q_v(\bm{O}_v, \bm{\alpha}_v|\bm{\theta}^Q_v)|_{\bm{O}_v=\bm{O}_v^{(i)}, \bm{\alpha}_v=\bm{\pi}_v(\bm{O}_v^{(i)})} \]  
\begin{equation}
\nabla_{\bm{\theta}_v^\pi} \mathcal{J} \approx \frac{1}{N} \sum_i J_a \cdot J_\theta
\label{eq:ddpg_grad} 
\end{equation}
where
\begin{equation}
\begin{cases}
J_a = \nabla_{\bm{\alpha}_v} Q_v(\bm{O}_v, \bm{\alpha}_v|\bm{\theta}^Q_v)|_{\bm{O}_v=\bm{O}_v^{(i)}, \bm{\alpha}_v=\bm{\pi}_v(\bm{O}_v^{(i)})} \\
J_\theta = \nabla_{\bm{\theta}_v^\pi} \bm{\pi}(\bm{O}_v|\bm{\theta}_v^\pi)|_{\bm{O}_v^{(i)}}
\end{cases}
\end{equation}
\ENDFOR
\ENDFOR
\RETURN Player $v$'s policy network, $\bm{\pi}_{v}(\bm{O}_v|\bm{\theta}_v^\pi)$.
\end{algorithmic}  
\end{algorithm} 

\subsection{Computational Cost}
\label{sec:execution_cost}
\input{content/execution_cost}

%% file: content/execution_cost.tex
\begin{figure}[h]
\centering
\begin{tabular}{cc}
  \includegraphics[width=0.22\textwidth]{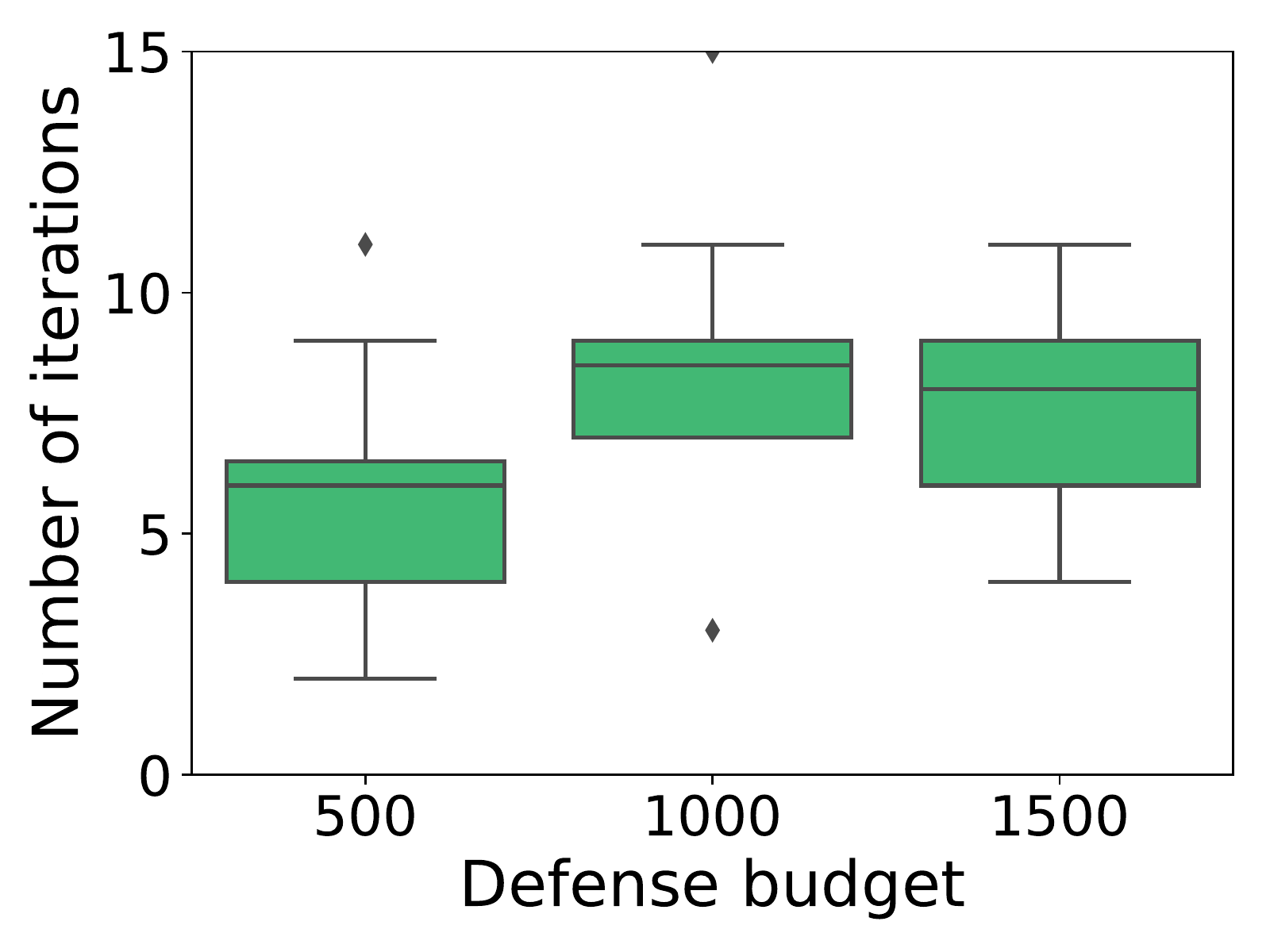} &
 \includegraphics[width=0.22\textwidth]{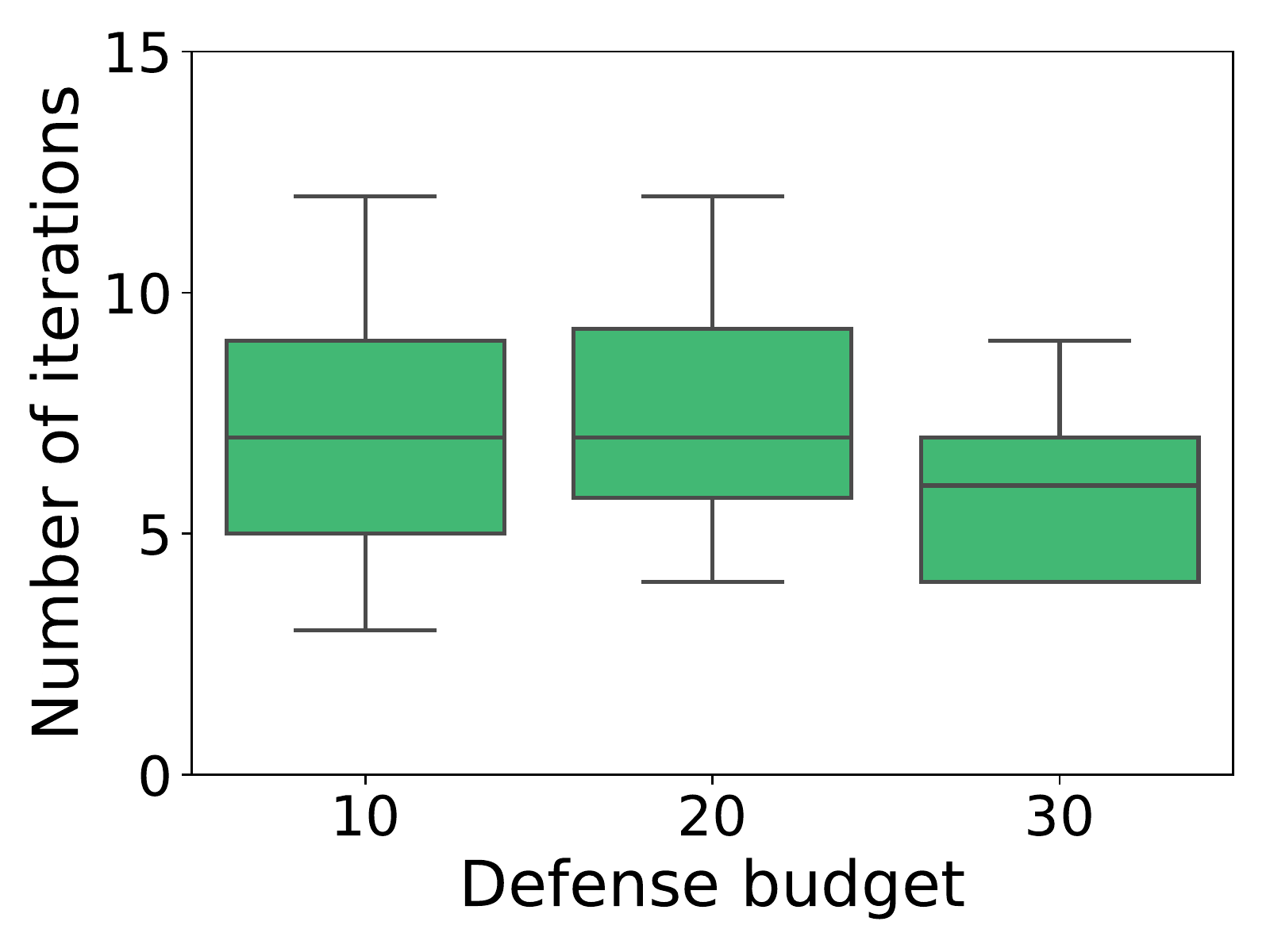}\\
%\multicolumn{2}{c}{\includegraphics[width=0.22\textwidth]{figures/execution_time}}
\end{tabular}
	\caption{Computational cost. Left: Number of double oracle iterations in intrusion detection with adv.\ budget=120.
	Right: Number of double oracle iterations in fraud detection with adv.\ budget=2.}
	\label{fig:execution_time}
\end{figure}

Figure~\ref{fig:execution_time} presents our evaluation of the computational cost of the proposed alert prioritization approach.
The results show that the double oracle algorithm can converge very fast in practice, with fewer than 15 iterations in most cases; indeed, in the vast majority of instances we need fewer than 10 iterations.

Another interesting observation is non-monotonicity of convergence time (in terms of iterations) as we increase the defense budget.
In the IDS setting, for example, increasing the defense budget increases the number of iterations when we go from a budget of 500 to 1000, but the computational cost remains stable as we further increase the budget to 1500.
In contrast, in the fraud detection case study, increasing the budget from 10 to 20 has little impact on the number of iterations, but further increasing it to 30 actually \emph{reduces} the number of iterations necessary for convergence.
To understand this phenomenon, note that increasing the defender's budget has two opposing effects: on the one hand, the search space for the defender increases significantly, but on the other hand, it may become much easier to compute a near-optimal defense with a larger budget (for example, with a large enough budget, we can almost always inspect all alerts).

%Moreover, it suggests that the computational cost of the double oracle method is affected by the defense budget.
%Specifically, larger defense budgets for intrusion detection lead to around $50\%$ more iterations before the double oracle algorithm stops.
%In contrast, a large defense budget results in $20\%$ fewer iterations in fraud detection.
%The main reason for this difference is that the state and action spaces in intrusion detection are much larger than those in fraud detection.
%Consequently, a larger defense budget makes the learning task more complex in intrusion detection; while in fraud detection, allocating the defense budget becomes easier as the defender obtains a budget that is adequate for investigating alerts.

%Figure~\ref{fig:execution_time} (bottom) shows the execution time of the proposed approach in both case studies.
%It can be seen that prioritizing alerts took longer in intrusion detection, about $25\%$ higher compared to fraud detection, due to its higher complexity.
%Moreover, we observed that updating the utility matrix of the double oracle accounts for more than $70\%$ of the execution time in both cases, which indicates that the update is the bottleneck in terms of computational performance.